\documentclass[authoryear,preprint,12pt]{elsarticle}

\usepackage{graphics}

\usepackage{amssymb}

\usepackage{amsthm}

\usepackage{amsmath}
\usepackage[authoryear]{natbib}

\newcommand{\bw}{\mathbf{w}}
\newcommand{\bx}{\mathbf{x}}

\newcommand{\bT}{\mathbf{T}}
\newcommand{\bz}{\mathbf{z}}

\newcommand{\bsr}{\boldsymbol{r}}
\newcommand{\bst}{\boldsymbol{t}}
\newcommand{\bsv}{\boldsymbol{v}}
\newcommand{\bsw}{\boldsymbol{w}}
\newcommand{\bsx}{\boldsymbol{x}}

\newcommand{\bsy}{\boldsymbol{y}}
\newcommand{\bW}{\boldsymbol{W}}

\newcommand{\bsmu}{\boldsymbol{\mu}}

\newcommand{\bbeta}{\boldsymbol{\beta}}
\newcommand{\bgamma}{\boldsymbol{\gamma}}
\newcommand{\btheta}{\boldsymbol{\theta}}
\newcommand{\bTheta}{\boldsymbol{\Theta}}
\newcommand{\bpsi}{\boldsymbol{\psi}}
\newcommand{\bPsi}{\boldsymbol{\Psi}}

\newcommand{\IR}{I\!\!R}
\journal{Neural Networks}

\begin{document}

\begin{frontmatter}

\title{Time series modeling by a regression approach based on a latent process}

\author[inrets,utc]{Faicel Chamroukhi\corref{cor1}}

\author[inrets]{Allou Sam\'{e}}
\author[utc]{G\'erard Govaert}
\author[inrets]{Patrice Aknin}
\address[inrets]{French National Institute for Transport and Safety Research (INRETS)\\
Laboratory of New Technologies (LTN)\\2 Rue de la Butte Verte, \\ 93166 Noisy-le-Grand Cedex (France)\fnref{label1}}
\address[utc]{Compi\`{e}gne University of Technology \\HEUDIASYC Laboratory, UMR CNRS 6599 \\
BP 20529, 60205 Compi\`{e}gne Cedex (France)}
\cortext[cor1]{Corresponding author:\\Faicel Chamroukhi\\ INRETS, 2
Rue de la Butte Verte,\\  93166 Noisy-le-Grand Cedex,  France\\Tel:
+33(1) 45 92 56 46\\Fax: +33(1) 45 92 55 01}
\ead{faicel.chamroukhi@inrets.fr}

\begin{abstract}
Time series are used in many domains including finance, engineering, economics and bioinformatics generally to represent the change of a measurement over time. Modeling techniques may then be used to give a synthetic representation of such data. A new approach for time series modeling is proposed in this paper. It consists of a regression model incorporating a discrete hidden logistic process allowing for activating smoothly or abruptly different polynomial regression models. The model parameters are estimated by the maximum likelihood method performed by a dedicated Expectation Maximization (EM) algorithm. The M step of the EM algorithm uses a multi-class Iterative Reweighted Least-Squares (IRLS) algorithm to estimate  the hidden process parameters. To evaluate the proposed approach, an experimental study on simulated data and real world data was performed using two alternative approaches: a heteroskedastic piecewise regression model using a global optimization algorithm based on dynamic programming, and a  Hidden Markov Regression Model whose parameters are estimated by the Baum-Welch algorithm. Finally, in the context of the remote monitoring of components of the French railway infrastructure, and more particularly the switch mechanism, the proposed approach has been applied to modeling and classifying time series representing the condition measurements acquired during switch operations.
\end{abstract}


\end{frontmatter}

\section{Introduction}
\label{sec: introduction}

\begin{figure}[htbp]
  \centerline{\includegraphics[height=5.5cm,width=6.5cm]{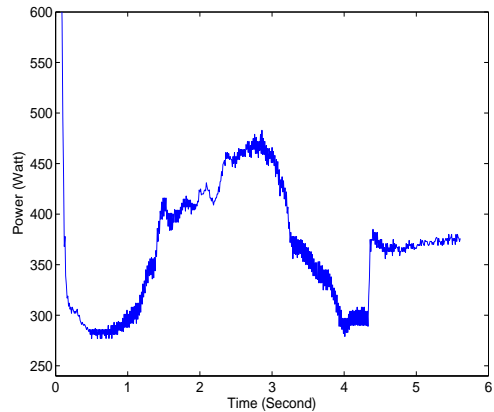}}
  \caption{A signal showing the electrical power consumed during a switch operation.}
  \label{signal_intro}
 \end{figure}

Time series occur in many domains including finance, engineering, economics, bioinformatics, and they generally represent the change of a measurement over time. Modeling techniques may then be used to give a synthetic representation of such data. This work relates to the diagnosis of the French railway switches (or points) which enable trains to be guided from one track to another at a railway junction. For this purpose, condition measurements acquired during switch operations are classified into predefined classes. Each measurement represents the electrical power consumed during a switch operation (see Fig. \ref{signal_intro}). 

The diagnosis task was performed by means of a two-step process:
feature extraction from the switch operation signals and the
implementation of a supervised learning algorithm to learn the
parameters of the operating classes of the switch mechanism. In this
paper we propose a new method for modeling switch operation signals.

Switch operations signals can be seen as time series presenting non-linearities and various changes in regime. In a context of this type, basic parametric methods based on linear or polynomial regression 
are not adapted. A piecewise regression model may be used as an alternative \citep{McGee,brailovsky,ferrari1}. Piecewise polynomial regression is a parametrization and segmentation method that partitions the data into $K$ segments, each segment being characterized by its mean polynomial curve and its variance. For this type of modeling, the parameters estimation can be exactly performed using dynamic programming algorithm \citep{bellman} such as Fisher's algorithm \citep{fisher}. This algorithm optimizes an additive  cost function over all the segments of the time series \citep{yveslechevalier90,brailovsky}. However, it is well-known that dynamic programming procedures are computationally expensive. An iterative algorithm can be derived to improve the running time of Fisher's algorithm as proposed by \citet{sameSFC2007}. This  approach iteratively estimates  the regression model parameters and the partition of the time series. The standard piecewise regression model usually assumes that noise variance is uniform in all the segments (homoskedastic model) \citep{brailovsky,ferrari1,ferrari2,sameSFC2007}. However, in this paper we shall consider a  heteroskedastic piecewise polynomial regression model. Another alternative approach is to use a Hidden Markov Regression
Model \citep{fridman} whose parameters are estimated by the
Baum-Welch algorithm \citep{BaumWelch}. However the piecewise and
Hidden Markov Regression approaches are more adapted for modeling
time series presenting abrupt changes and may be less efficient for
time series including regimes with smooth transitions.

The method we propose for time series modeling is based on a
specific regression model incorporating a discrete hidden process
allowing for abrupt or smooth transitions between different
regression models. This approach is related to the switching
regression model introduced by \citet{quandt} and is very linked to
the Mixture of Experts (ME) model developed  by \citet{jordanHME} by
the using of a time-dependent logistic transition function. The ME
model, as discussed in \citep{waterhouse}, uses a conditional
mixture modeling where the model parameters are estimated by the
Expectation Maximization (EM) algorithm \citep{dlr, mclachlanEM}.
Once the model parameters of the proposed regression model with
hidden process are estimated, they are used as the feature vector
for each signal. The parameters of the different operating classes
(no defect, minor defect and critical defect) are then learnt from a
labelled collection of signals using Mixture Discriminant Analysis
(MDA) \citep{tibshiraniMDA}. Based on the operating classes
parameters, a new signal is classified by using the Maximum A
Posteriori (MAP) rule. The good performance of the proposed approach
has been demonstrated by an experimental study carried out on real
measured signals covering a wide range of defects.

This paper is organized as follows. Section 2 provides an account of
the heteroskedastic piecewise polynomial regression model,  and the
parameter estimation technique this uses based on a dynamic
programming procedure. Section 3 presents the Hidden Markov
Regression Model whose parameters are estimated by the Expectation
Maximization Baum-Welch algorithm. Section 4 introduces the proposed
model and describes parameters estimation by means of the EM
algorithm. Section 5 deals with the experimental study that assesses
the performance of the proposed approach in terms of signal modeling
and section 6 describes the application of the proposed technique to
switch operation signals modeling and classification.

\section{The piecewise polynomial regression model}
\label{sec: piecewise polynomial regression model}

Let $\bx=(x_1,\ldots,x_n)$ be $n$ real observations of a signal or a time series where $x_i$ is observed at time $t_i$. The piecewise polynomial regression model assumes that the time series incorporates $K$ polynomial regimes on $K$ intervals whose bounds indexes can be denoted by $\bgamma = (\gamma_1,\ldots,\gamma_{K+1})$ with $\gamma_1=0$ and $\gamma_{K+1}=n$. This defines a partition of the time series into $K$ polynomial segments $(\bsx_1,\ldots,\bsx_K)$ of lengths $n_1,\ldots,n_K$ where $\bsx_k = \{x_i|i\in I_k\}$ is the set of elements in segment $k$ whose indexes are $I_k = (\gamma_{k},\gamma_{k+1}]$.

Standard polynomial regression models are homoskedastic models as they assume that the different polynomial regression models have the same noise variance. In our case we shall consider the more general framework of a heteroskedastic model which allows the noise level to vary between the different polynomial regression models. It can be defined as follows:
\begin{equation}
\forall i=1,\ldots,n, \quad x_i = \bbeta^T_k\bsr_{i} +\sigma_k\varepsilon_{i} \quad ; \quad\varepsilon_{i} \sim \mathcal{N}(0,1) ,
\label{eq.piecewise regression model}
\end{equation}
where $k$ satisfies $i \in I_k$, $\bbeta_k$ is the $(p+1)$-dimensional coefficients vector of a $p$ degree polynomial associated with the $k^{th}$ segment with $k \in \{1,\ldots,K\}$, \linebreak $\bsr_{i}=(1,t_i,t_i^2\ldots,t_i^p)^T$ is the time dependent $(p+1)$-dimensional covariate vector associated to the parameter $\bbeta_{k}$ and the $\varepsilon_{i}$ are independent random variables with a standard Gaussian distribution representing the additive noise in each segment $k$.

\subsection{Maximum likelihood estimation for the piecewise polynomial regression model}

With this model,  the parameters can be denoted by $(\bpsi,\bgamma)$
where \linebreak
$\bpsi=(\bbeta_1,\ldots,\bbeta_K,\sigma_1^2,\ldots,\sigma_K^2)$ is
the set of polynomial coefficients and noise variances, and
$\bgamma=(\gamma_{1},\ldots,\gamma_{K+1})$ is the set of transition
points. Parameter estimation is performed by maximum likelihood. We
assume a conditional independence of the data. Thus, according to
the model defined by equation (\ref{eq.piecewise regression model}),
it can be proved that within each segment $k$, $x_i$ has a Gaussian
distribution with mean $\bbeta_k^T\bsr_i$ and variance $\sigma_k^2$,
and therefore, the log-likelihood of the parameter vector
$(\bpsi,\bgamma)$ characterizing the piecewise regression model is
the sum of the local log-likelihoods over the $K$ segments that can
be written as follows
\begin{eqnarray}
L(\bpsi,\bgamma;\bx) &=& \log p(\bx;\bpsi,\bgamma) \nonumber \\
&=&\sum_{k=1}^K \sum_{i\in I_k} \log \mathcal{N}\left(x_i;\bbeta_k^T\bsr_i,\sigma_k^2\right).
\end{eqnarray}

Maximizing this log-likelihood is equivalent to minimizing with
respect to $\bpsi$ and $\bgamma$ the criterion
\begin{eqnarray}
J(\bpsi,\bgamma) &=&\sum_{k=1}^K
\left[\frac{1}{\sigma_k^2}\sum_{i\in
I_k}\left(x_i-\bbeta_k^{T}\bsr_i \right)^2 + n_k \log \sigma_k^2
\right], \label{eq.picewise_reg criterion J}
\end{eqnarray}
where $n_k$ is the number of elements in segment $k$.

Since the criterion $J$ is additive over the $K$ segments, the
Fisher's algorithm \citep{fisher,yveslechevalier90},
which consists in a dynamic programming procedure \citep{bellman,brailovsky},
can be used to perform the global minimization. This dynamic
procedure has a time complexity of $O(Kp^2n^2)$ which can be
computationally expensive for large sample sizes.

\subsection{Time series approximation and segmentation with the piecewise regression model}

Once the parameters have been estimated, a segmentation of the time
series, equivalently represented by the classes vector
$\hat{\bz}=(\hat{z}_1,\ldots,\hat{z}_n)$,  where
$\hat{z}_i\in \{1,\ldots,K \}$, can be derived by setting
$\hat{z}_i=k \mbox{ if } i \in
(\hat{\gamma}_k;\hat{\gamma}_{k+1}]$, the parameters
$(\hat{\bpsi},\hat{\bgamma})$ being the parameters provided
by the dynamic programming procedure.

An approximation of the time series is then given by $\hat{x}_i
= \sum_{k=1}^K \hat{z}_{ik} \hat{\bbeta}^T_k \bsr_i$, where
$\hat{z}_{ik}=1$ if $\hat{z}_i=k$ and $\hat{z}_{ik}=0$
otherwise. The vectorial formulation of the approximated time series
$\hat{\bx}$ can be written as:
\begin{equation}
\hat{\bx} = \sum^{K}_{k=1} \hat{Z}_{k} \bT \hat{\bbeta}_{k},
\end{equation}
where $\hat{Z}_{k}$ is a diagonal matrix whose diagonal elements are
$(\hat{z}_{1k} ,\ldots,\hat{z}_{nk} )$, and
$$\bT=\left[\begin{array}{ccccc}
1&t_{1}&t_{1}^2&\ldots&t_{1}^p\\
1&t_{2}&t_{2}^2&\ldots&t_{2}^p \\
\vdots&\vdots&\vdots&\vdots&\vdots\\
1&t_{n}&t_{n}^2&\ldots&t_{n}^p\end{array}\right]$$ is the $[n\times
(p+1)]$ regression matrix.

\section{The Hidden Markov Regression Model }
\label{sec: Hidden Markov Model Regression}

This section recalls the Hidden Markov Regression Model (HMRM)
\linebreak \citep{fridman}. Owing to the fact that the real signals
we want to model consist of successive phases, order constraints are
assumed
for the hidden states in the HMRM.

\subsection{A general description of Hidden Markov Regression Models}
\label{ssec: HMM regression model}

In a  Hidden Markov Regression Model, the time series is represented
as a sequence of observed variables $\bx = (x_1,\ldots,x_n)$, where
$x_i$ is observed at time $t_i$ and assumed to be generated by the
following regression model \citep{fridman}:
\begin{equation}
\forall i=1,\ldots,n,  \quad  x_{i} = \bbeta^T_{z_i}\bsr_{i} + \sigma_{z_i}\varepsilon_{i} \quad ; \quad\varepsilon_{i} \sim \mathcal{N}(0,1) ,
\label{eq.HMM regression model}
\end{equation}
{where ${z_{i}}$ is a discrete hidden variable taking its values in
the set $\{1,\ldots,K\}$.}

The HMRM assumes that the hidden variable $\bz=(z_1,\ldots,z_n)$ is
a homogeneous Markov chain where the variable $z_i$ controls the
switching from one polynomial regression model to another of $K$
models at each time $t_i$. The distribution of the latent sequence
$\bz=(z_1,\ldots,z_n)$ is defined as:
\begin{eqnarray}
p(\bz;\pi,A) &=&p(z_1;\pi) \prod_{i=2}^{n}p(z_i \vert z_{i-1};A) \nonumber \\
&=& \prod_{k=1}^{K}\pi_{k}^{z_{1k}} \prod_{i=2}^{n} \prod_{k=1}^{K} \left[\prod_{\ell=1}^{K} {A_{\ell k}}^{z_{(i-1)\ell}}\right]^{z_{ik}},
\label{eq.HMM_process}
\end{eqnarray}
where
\begin{itemize}
\item $\pi =(\pi_1,\ldots,\pi_K)$  is the initial distribution of $z_i$, with $\pi_{k} = p(z_1 = k)$ for $k\in\{1,\ldots,K\}$;
\item $A=(A_{\ell k})_{1\leq \ell,k\leq K}$ where $A_{\ell k}=p(z_i=k|z_{i-1}=\ell)$ is the matrix of transition probabilities;
\item $z_{ik}=1$ if $z_{i}=k$ (i.e if $x_i$ is generated by the $k^{th}$ regression model) and $z_{ik}=0$ otherwise.
\end{itemize}

\subsection{Parameter estimation of the Hidden Markov Regression Model}
\label{sec: parameter estimation of HMM regression}

From the model defined by equation (\ref{eq.HMM regression model}), it
can be proved that, conditionally on  a regression model $k$
($z_i=k$), $x_i$ has a Gaussian distribution with mean
$\bbeta_k^T\bsr_i$ and variance $\sigma_k^2$. Thus, the HMRM is
parameterized by the parameter vector
$\bPsi=(\pi,A,\bbeta_1,\ldots,\bbeta_K,\sigma^2_1,\ldots,\sigma^2_K)$.
The parameter vector $\bPsi$ is estimated by the maximum likelihood
method. The log-likelihood to be maximized in this case is written
as:
\begin{eqnarray}
L(\bPsi;\bx)&=&\log p(\bx;\bPsi) \nonumber\\
&=& \log\sum_{\bz} p(z_1;\pi)\prod_{i=2}^{n}p(z_i|z_{i-1};A)\prod_{i=1}^{n}\mathcal{N}(x_i;\bbeta^T_{z_i},\sigma_{z_i}^2).
\label{HMM_regression_logik}
\end{eqnarray}
Since this log-likelihood can not be maximized directly, this is done
by the EM algorithm \citep{dlr}, which is known as the Baum-Welch
algorithm \citep{BaumWelch} in the context of HMMs. It can easily be
verified that, in a regression context, the Baum-Welch algorithm has
a time complexity of $O(IKp^2n)$, where $I$ is the number of
iterations of the algorithm.

\subsection{A HMRM with order constraints}
\label{constraint transmat}

Since the switch operation signals we aim to model consist of
successive phases, we impose the following constraints on the
transition probabilities:
\begin{equation}
p(z_i=k|z_{i-1}= \ell) = 0 \quad \mbox{if} \quad k<\ell \; ,
\label{HMM_contraint1}
\end{equation}
and
\begin{equation}
p(z_i=k|z_{i-1}= \ell) = 0 \quad \mbox{if} \quad k>\ell + 1.
\label{HMM_contraint2}
\end{equation}

These constraints imply that no transitions are allowed for the phases whose indices are lower that the current phase (equation \ref{HMM_contraint1}) and no jumps of more than one state are possible (equation \ref{HMM_contraint2}). This constrained model is a  particular case of the well known
left-right model \citep{rabiner}.

\subsection{Time series approximation and segmentation with the HMRM}
\label{sse: estimation of the denoised signal for HMM regression}

To approximate the time series, at each time $t_i$ we combine the
different regression models using the filtering probabilities
denoted by $\omega_{ik}$ for the $k^{th}$ regression model. The
filtering probability is the probability \linebreak
$\omega_{ik}=p(z_i=k|x_1,\ldots,x_{i};\bPsi)$ that $x_i$ will be
generated by the regression model $k$ given the observations
$(x_1,\ldots,x_i)$ that occur until time $t_i$. It can be computed
using the so-called ``forward" probabilities \citep{rabiner}. Thus, the
filtered time series $\hat{\bx}=(\hat{x}_1,\ldots,\hat{x}_n)$, which
is common way to approximate the time series $\bx$, is given by:
\begin{eqnarray}
\hat{x}_i&=& \sum_{k=1}^{K} \hat{\omega}_{ik}\hat{\bbeta}^T_k
\bsr_{i} \; ; \; i=1,\ldots,n,
\end{eqnarray}
where $\hat{\bPsi} =
(\hat{\pi},\hat{A},\hat{\bbeta}_1,\ldots,\hat{\bbeta}_K,\hat{\sigma}^2_1,\ldots,\hat{\sigma}^2_K)$
and $\hat{\omega}_{ik}$ are respectively the parameter vector and
the filtering probability obtained using the EM (Baum-Welch)
algorithm. The vectorial formulation of the approximated time series
$\hat{\bx}$ can be written as:
\begin{equation}
\hat{\bx} = \sum^{K}_{k=1} \hat{\mathcal{W}}_{k} \bT \hat{\bbeta}_{k},
\label{eq. signal approximation in HMRM}
\end{equation}
where $\hat{\mathcal{W}}_{k}$ is a diagonal matrix whose diagonal elements are $(\hat{\omega}_{1k} ,\ldots,\hat{\omega}_{nk} )$, and $\bT$ is the $[n\times (p+1)]$ regression matrix. This approximation will be taken as the denoised signal.

On the other hand, a segmentation of the time series can be deduced
by computing the label $\hat{z_i}$ of $x_i$ using the Maximum A
Posteriori (MAP) rule as follows:
\begin{equation}
\hat{z_i} = \arg \max \limits_{\substack {1\leq k\leq K}}
\hat{\tau}_{ik} \; ; \;  \forall i=1,\ldots ,n,
\end{equation}
where $\tau_{ik} = p(z_i=k|\bx;\bPsi)$ is the posterior probability
that $x_i$ originates from the $k^{th}$ regression model. Notice
that $\tau_{ik}$ can be computed using the ``forward" and
``backward" probabilities \citep{rabiner}.

\section{The proposed regression model with a hidden logistic process}
\label{sec: regression model}

The proposed regression model introduced in this section is defined,
as for the HMRM model, by equation (\ref{eq.HMM regression model}),
where a logistic process is used to model the hidden sequence
$\bz=(z_1,\ldots,z_n)$.

\subsection{The hidden logistic process}
\label{ssec: process}

This section defines the probability distribution of the process \linebreak $\bz=(z_1,\ldots,z_n)$ that allows the switching from one regression model to another.

The proposed hidden logistic process assumes that the variables $z_i$, given the vector $\bst=(t_1,\ldots,t_n)$, are generated independently according to the multinomial distribution {\small$\mathcal{M}(1,\pi_{i1}(\bw),\ldots,\pi_{iK}(\bw))$}, where
\begin{equation}
\pi_{ik}(\bw)= p(z_i=k;\bw)=\frac{\exp{(\bsw_k^T\bsv_{i})}}{\sum_{\ell=1}^K\exp{(\bsw_{\ell}^T \bsv_{i})}},
\label{multinomial logit}
\end{equation}
is the logistic transformation of a linear function of the time-dependent covariate $\bsv_i=(1,t_i,t_{i}^2,\ldots,t_i^q)^T$, $\bsw_{k}=(\bsw_{k0},\ldots,\bsw_{kq})^T$ is the $(q+1)$-dimensional coefficients vector associated with the covariate $\bsv_i$ and $\bw = (\bsw_1,\ldots,\bsw_K)$. Thus, given the vector $\bst=(t_1,\ldots,t_n)$, the distribution of $\bz$ can be written as:
\begin{equation}
p(\bz;\bw)=\prod_{i=1}^n \prod_{k=1}^K \left(\frac{\exp{(\bsw_{k}^T\bsv_{i})}}{\sum_{\ell=1}^K\exp{(\bsw_{\ell}^T \bsv_{i})}}\right)^{z_{ik}} ,
\label{eq.hidden logistic process}
\end{equation}
where $z_{ik} = 1$ if $z_i=k$ i.e when $x_i$ is generated by the $k^{th}$ regression model, and $0$ otherwise.

The relevance of the logistic transformation in terms of flexibility of transition can be illustrated through simple examples with $K=2$ components. In this case, only the probability $\pi_{i1}(\bw)= \frac{exp(\bsw^T_1 \bsv_i)}{1+exp(\bsw^T_1 \bsv_i)}$ should be described, since $\pi_{i2}(\bw)=1-\pi_{i1}(\bw)$. The first example is designed to show the effect of the dimension $q$ of $\bsw_k$ on the temporal variation of the probabilities $\pi_{ik}$. We consider different values of the dimension $q$ ($q=0,1,2$) of $\bsw_k$.

\begin{figure*}[htbp]
\centering
\begin{tabular}{ccc}
\includegraphics[height=4cm,width=4.2cm]{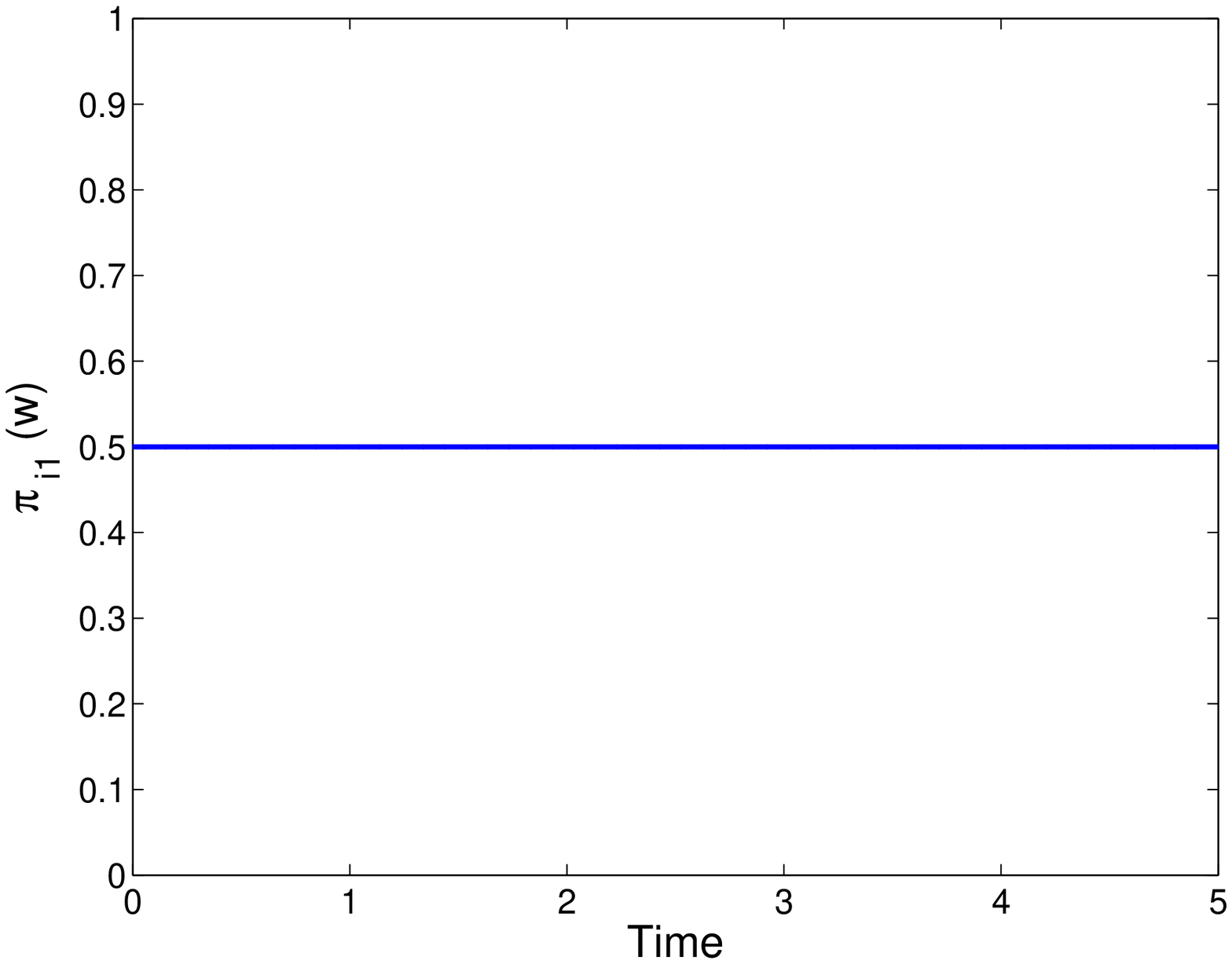}&

\includegraphics[height=4cm,width=4.2cm]{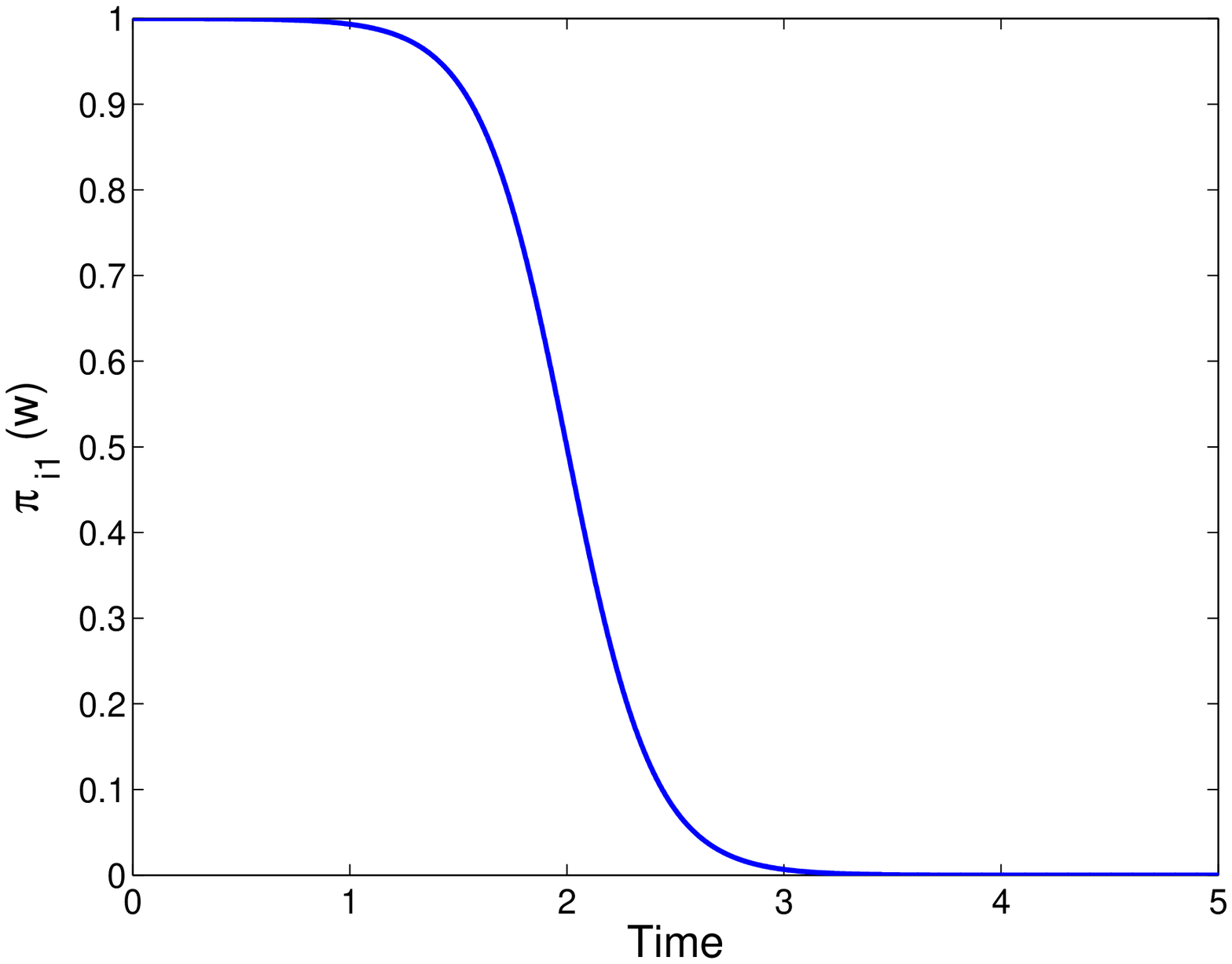}&

\includegraphics[height=4cm,width=4.2cm]{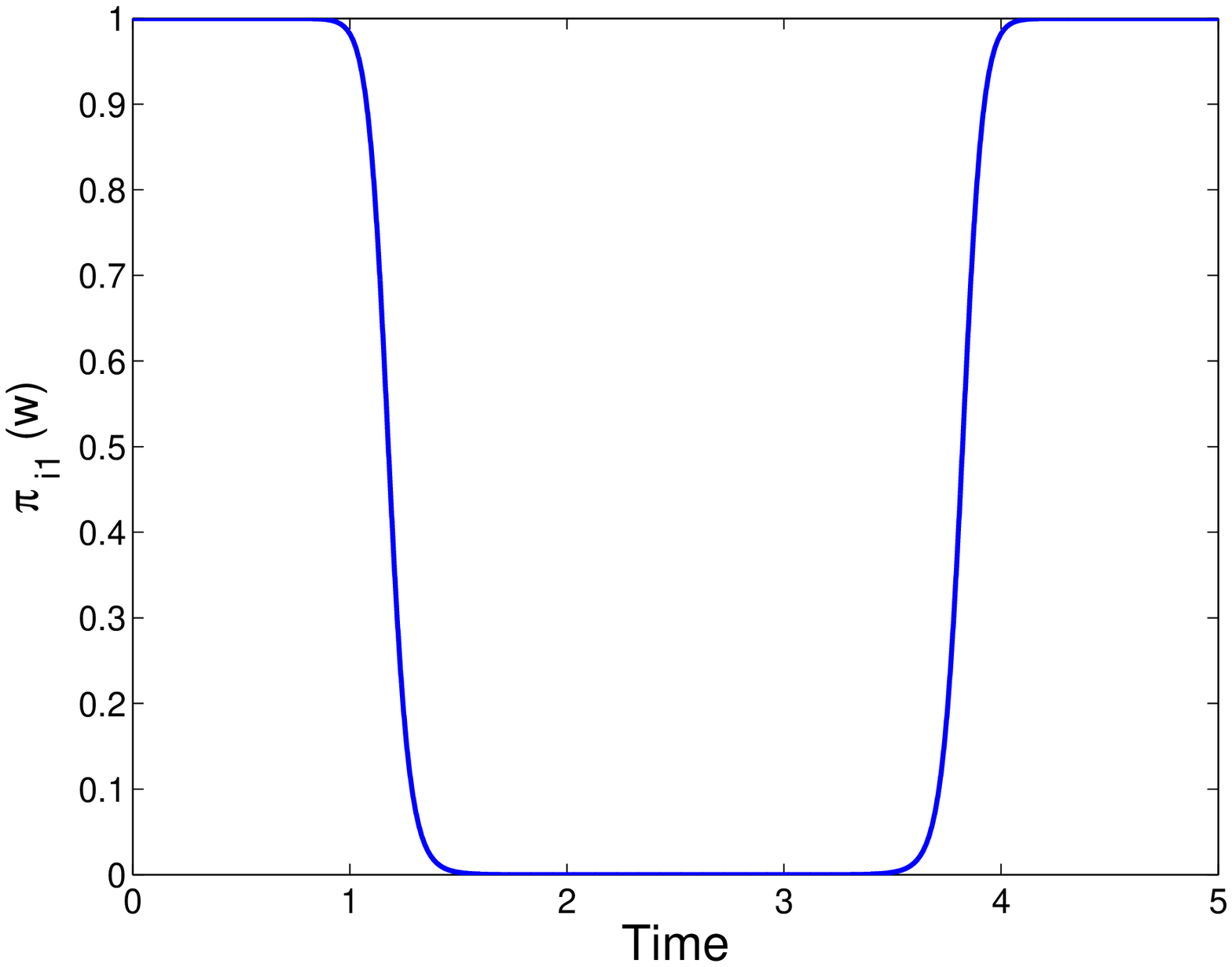}\\
\small{(a)}&\small{(b)}&\small{(c)}
\end{tabular}
\caption{Variation of $\pi_{i1}(\bw)$ over time for different values of the dimension $q$ of $\bsw_1$, for $K = 2$ and (a) $q=0$ and $\bsw_1=0$, (b) $q=1$ and $\bsw_1 = (10,-5)^T$ and (c) $q= 2$ and $\bsw_1 = (-10,-20,-4)^T$.}
\label{logistic_function_k=2_q=012}
\end{figure*}

As shown in Fig. \ref{logistic_function_k=2_q=012}, the dimension
$q$ controls the number of  temporal transitions of $\pi_{ik} $. In
fact, the larger the dimension of $\bsw_k$, the more complex the
temporal variation of $\pi_{ik}$. More particularly, if the goal is
to segment the signals into contiguous segments, the dimension $q$
of $\bsw_k$ must be set to $1$, what will be assumed in the
following.

\begin{figure*}[htbp]
\centering
\begin{tabular}{cc}
\includegraphics[height=5.5cm,width=6.5cm]{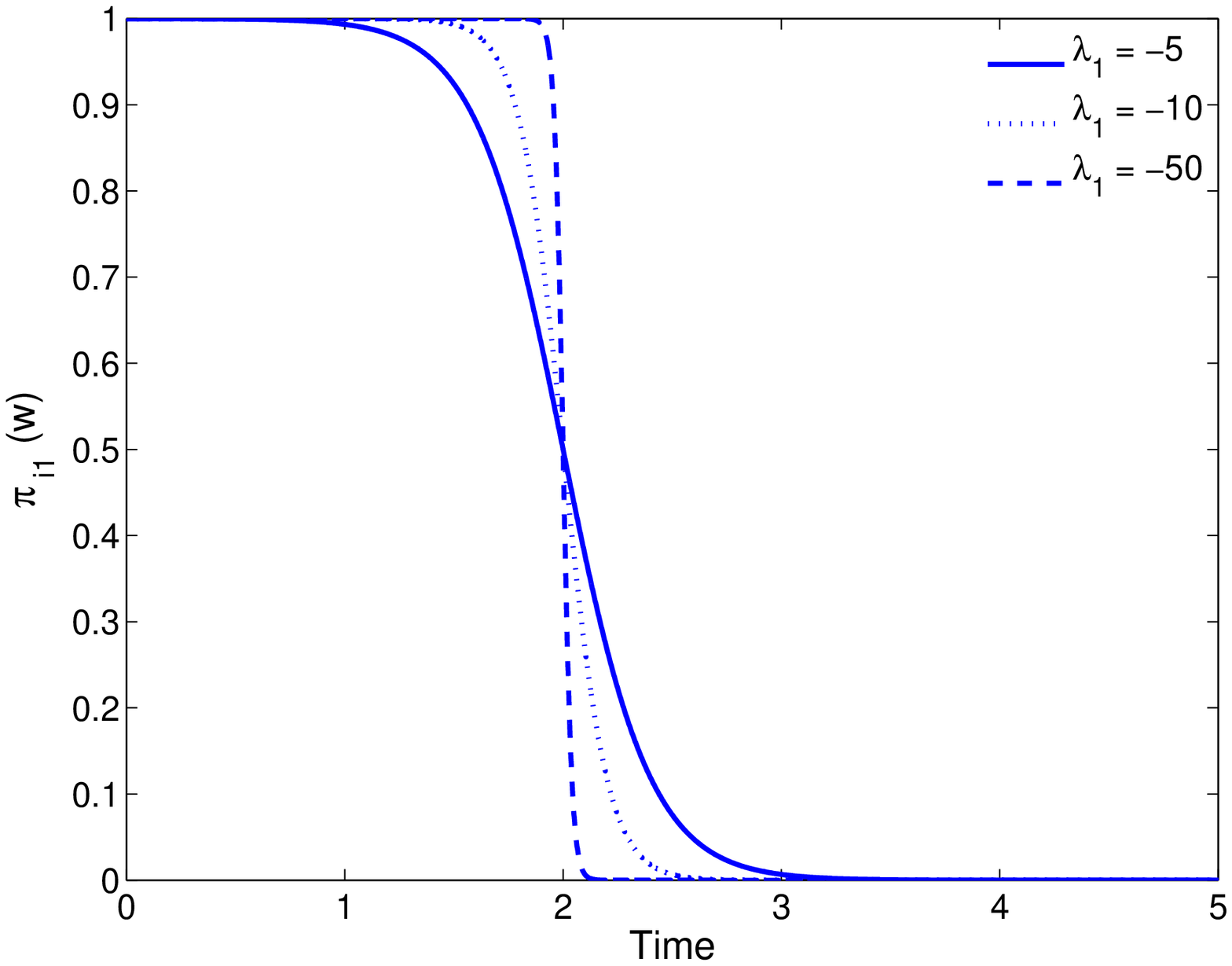} &
\includegraphics[height=5.5cm,width=6.5cm]{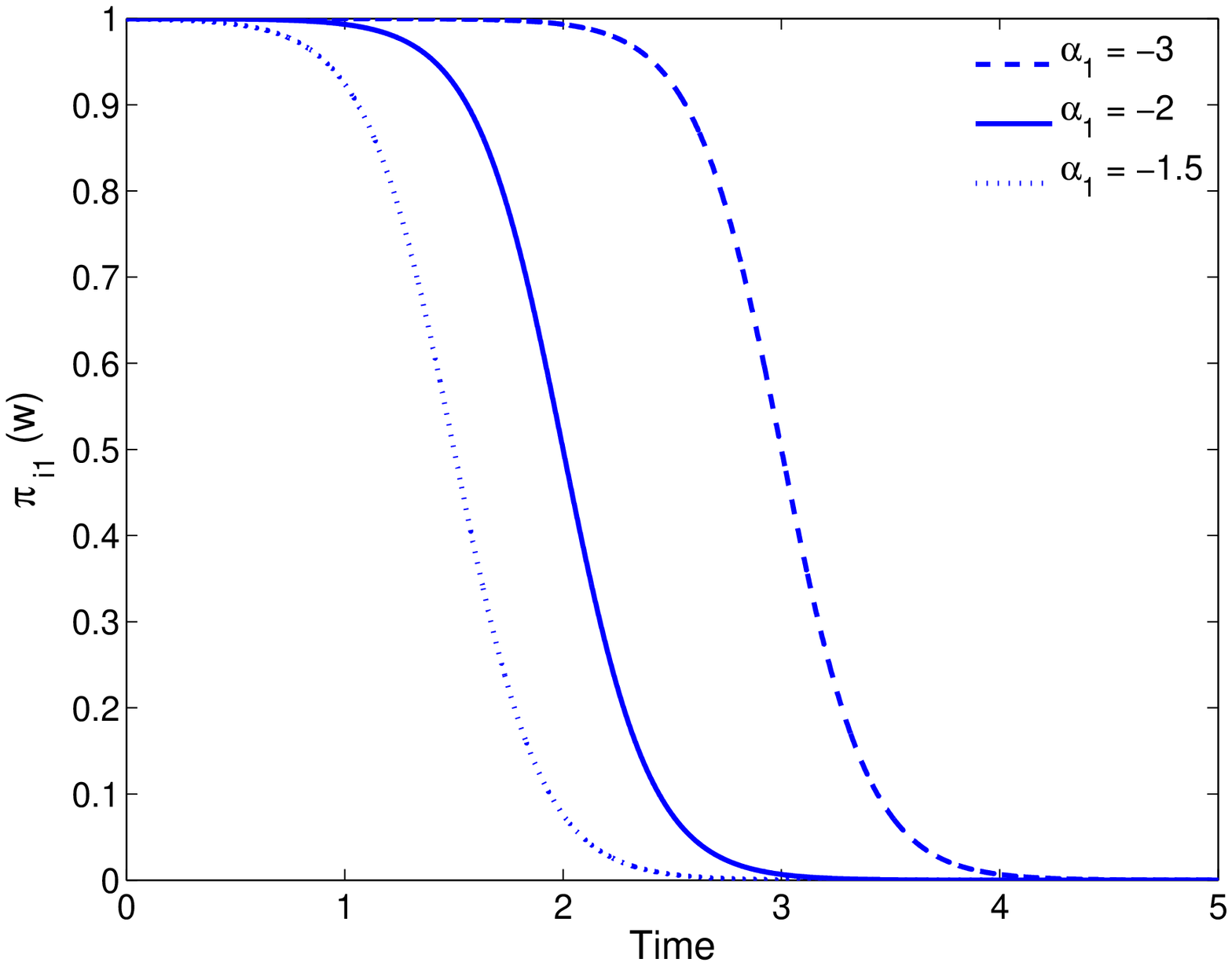}\\
\small{(a)} & \small{(b)}
\end{tabular}
\caption{Variation of $\pi_{i1}(\bw)$ over time for a dimension $q=1$ of $\bsw_1$ and (a) different values of $\lambda_1 = \bsw_{11}$ with $\alpha_1 = -2$ and (b) different values of $\alpha_1 = \frac{\bsw_{10}}{\bsw_{11}}$ with $\lambda_1=-5$.}
\label{logistic_function_k=2_q=1}
\end{figure*}

For a fixed dimension $q$ of the parameter $\bsw_k$, the variation of the proportions $\pi_{ik}(\bw)$ over time, in relation to the parameter  $\bsw_k$, is illustrated by an example of 2 classes with $q=1$. For this purpose, we use the parametrization $\bsw_k =\lambda_k (\alpha_k, 1)^T$ of $\bsw_k$, where $\lambda_k= \bsw_{k1}$ and $\alpha_k = \frac{\bsw_{k0}}{\bsw_{k1}} \cdot$ As shown in Fig. \ref{logistic_function_k=2_q=1} (a), the parameter $\lambda_k$ controls the quality of transitions between classes, the higher absolute value of $\lambda_k$, the more abrupt the transition between the $z_i$, while the parameter $\alpha_k$ controls the transition time point via the inflexion point of the curve (see Fig. \ref{logistic_function_k=2_q=1} (b)).  

In this particular regression model, the variable $z_i$ controls the switching from one regression model to another of $K$ regression models at each time $t_i$. Therefore, unlike basic polynomial regression models, which assume uniform regression parameters over time, the proposed model permits the polynomial coefficients to vary over time by switching from one regression model to another.

\subsection{The generative model for signals}

The generative model that produces a signal from a fixed parameter \linebreak $\btheta=\{\bsw_k,\bbeta_k,\sigma^2_k ; k=1,\ldots,K\}$ consists of 2 steps:  
\begin{itemize}
\item generate the hidden process $\bz=(z_1,\ldots,z_n)$ according to the multinomial distribution $z_i \sim {\small \mathcal{M}(1,\pi_{i1}(\bw),\ldots,\pi_{iK}(\bw))}$,
\item  generate each observation $x_i$ according to the Gaussian distribution $\mathcal{N}(\cdot;\bbeta^{T}_{z_i}\bsr_i,\sigma^{2}_{z_i})$.
\end{itemize}

\subsection{Parameter estimation}
\label{ssec: parameter estimation HLP regression}

From the proposed model, it can be proved that, conditionally on a regression model $k$, $x_i$ is distributed according to a normal density with mean $\bbeta^T_k\bsr_{i}$ and variance $\sigma^2_k$. Thus, it can be proved that $x_{i}$ is distributed according to the normal mixture density
\begin{equation}
p(x_{i};\btheta)=\sum_{k=1}^K\pi_{ik}(\bw)\mathcal{N}\big(x_{i};\bbeta^T_k\bsr_{i},\sigma^2_k\big) ,
\label{melange}
\end{equation}
where $\btheta=(\bsw_{1},\ldots,\bsw_{K},\bbeta_1,\ldots,\bbeta_K,\sigma^2_1,\ldots,\sigma^2_K)$
is the parameter vector to be estimated. The parameter $\btheta$ is estimated by the maximum likelihood method. As in the classic regression models we assume that, given $\bst =(t_1,\ldots,t_n)$,  the $\varepsilon_i$ are independent. This also implies the independence of $x_i$ $(i=1,\ldots,n)$. The log-likelihood of $\btheta$ is then written as:
\begin{eqnarray}
L(\btheta;\bx)&=&\log \prod_{i=1}^np(x_i;\btheta)\nonumber\\
&=&\sum_{i=1}^{n}\log\sum_{k=1}^K \pi_{ik}(\bw)\mathcal{N}\big(x_{i};\bbeta^T_k\bsr_i,\sigma^2_k\big) .
\end{eqnarray}
Since the direct maximization of this likelihood is not
straightforward, it is maximized with the Expectation Maximization
(EM) algorithm \citep{dlr,mclachlanEM}.

\subsection{The dedicated EM algorithm}
\label{ssec. EM algortihm}

The proposed EM algorithm starts from an initial parameter $\btheta^{(0)}$ and alternates the two following steps until convergence:

\subsubsection{ E Step (Expectation) }

This step consists in computing the expectation of the complete log-likelihood $\log p(\bx,\bz;\btheta)$, given the observations and the current value $\btheta^{(m)}$ of the parameter $\btheta$ ($m$ being the current iteration):
\begin{eqnarray}
\!\!\!\! Q(\btheta,\btheta^{(m)})&\!\!\!\!=\!\!\!\!& E\left[\log p(\bx,\bz;\btheta)|\bx;\btheta^{(m)}\right]\nonumber\\
&\!\!\!\!=\!\!\!\! &  \sum_{i=1}^{n}\sum_{k=1}^K E (z_{ik}|x_i;\btheta^{(m)}) \log \left[\pi_{ik}(\bw)\mathcal{N}(x_{i};\bbeta^T_k\bsr_{i},\sigma_k^2)\right] \nonumber \\
&\!\!\!\!=\!\!\!\! &\sum_{i=1}^{n}\sum_{k=1}^K \tau^{(m)}_{ik}\log \left[\pi_{ik}(\bw)\mathcal{N} \left(x_{i};\bbeta^T_k\bsr_{i},\sigma^2_k \right)\right] \nonumber \\
&\!\!\!\!=\!\!\!\! &\sum_{i=1}^{n}\sum_{k=1}^K \tau^{(m)}_{ik}\log \pi_{ik}(\bw) + \sum_{i=1}^{n}\sum_{k=1}^K \tau^{(m)}_{ik}\log \mathcal{N} \left(x_{i};\bbeta^T_k\bsr_{i},\sigma^2_k \right),
\end{eqnarray}
where
\begin{eqnarray}
 \tau^{(m)}_{ik} &=& p(z_{ik}=1|x_i;\btheta^{(m)})
 =\frac{\pi_{ik}(\bw^{(m)})\mathcal{N}(x_{i};\bbeta^{T(m)}_k\bsr_{i},\sigma^{2(m)}_k)}
{\sum_{\ell=1}^K\pi_{i
\ell}(\bw^{(m)})\mathcal{N}(x_{i};\bbeta^{T(m)}_{\ell}\bsr_{i},\sigma^{2(m)}_{\ell})}\;
\label{eq.tik}
\end{eqnarray}
is the posterior probability that $x_i$ originates from the $k^{th}$ regression model.
\\As shown in the expression for $Q$, this step simply requires the computation of $\tau^{(m)}_{ik}$.

\subsubsection{ M step (Maximization) }

In this step, the value of the parameter $\btheta$ is updated by computing the parameter $\btheta^{(m+1)}$ maximizing the conditional expectation $Q$ with respect to $\btheta$. To perform this maximization, it can be observed that $Q$ is written as:
\begin{equation}
Q(\btheta,\btheta^{(m)})=Q_1(\bw)+\sum_{k=1}^K Q_2(\bbeta_k,\sigma^2_k),
\end{equation}
with
\begin{equation}
Q_1(\bw)=\sum_{i=1}^{n}\sum_{k=1}^K \tau^{(m)}_{ik}\log \pi_{ik}(\bw),
\end{equation}
and,
\begin{eqnarray}
Q_2(\bbeta_k,\sigma^2_k) &=& \sum_{i=1}^{n} \tau^{(m)}_{ik}\log \mathcal{N}\left(x_{i};\bbeta^T_k\bsr_{i},\sigma^2_k\right) \nonumber \\
&=& -\frac{1}{2} \left[\frac{1}{\sigma_k^2}\sum_{i=1}^{n} \tau^{(m)}_{ik}\left(x_{i}-\bbeta^T_k\bsr_{i}\right)^2 +n_k^{(m)}\log \sigma_k^2 \right] \nonumber \\
& &-\frac{n_k^{(m)}}{2}\log 2\pi\; ; \; k=1,\ldots,K ,
\end{eqnarray}
where $n_k^{(m)}=\sum_{i=1}^{n} \tau^{(m)}_{ik}$ can be interpreted
as the number of points of the component $k$ estimated at the
iteration $m$. Thus, the maximization of $Q$ can be performed by
separately maximizing $Q_1(\bw)$ with respect to $\bw$ and
$Q_2(\bbeta_k,\sigma^2_k)$ with respect to $(\bbeta_k,\sigma^2_k)$
for all $k=1,\ldots,K$. Maximizing  $Q_2$ with respect to $\bbeta_k$
consists in analytically solving a weighted least-squares problem.
The estimates are given by:
\begin{eqnarray}
{\bbeta}_k^{(m+1)} &=& \arg \min \limits_{\substack {\bbeta_k}} \sum_{i=1}^{n} \tau^{(m)}_{ik} ( x_i-\bbeta_k^{T}\bsr_i)^2 \nonumber \\
&=& (\bT^T\bW_k^{(m)}\bT)^{-1}\bT^T\bW_k^{(m)}\bx,
\label{estimation betaEM}
\end{eqnarray}
where $\bW_k^{(m)}$ is the $[n \times n]$ diagonal matrix of weights whose diagonal elements are $(\tau_{1k}^{(m)},\ldots,\tau_{nk}^{(m)})$ and $\bT$ is the $[n\times (p+1)]$ regression matrix.

Maximizing  $Q_2$ with respect to $\sigma_k^2$ provides the following updating formula:
\begin{eqnarray}
{\sigma}_k^{2(m+1)}&=& \arg \min \limits_{\substack{\sigma_k^2}} \left[\frac{1}{\sigma_k^2}\sum_{i=1}^{n} \tau^{(m)}_{ik}\left(x_{i}-\bbeta^{T(m+1)}_k\bsr_{i}\right)^2 +n_k^{(m)}\log \sigma_k^2 \right] \nonumber \\
&=& \frac{1}{n_k^{(m)}}\sum_{i=1}^{n} \tau^{(m)}_{ik} (x_i-{\bbeta}_k^{T(m+1)}\bsr_i)^2 .
\label{estimation sigmaEM}
\end{eqnarray}

The maximization of $Q_1$ with respect to $\bw$ is a multinomial
logistic regression problem weighted by $\tau^{(m)}_{ik}$ which we
solve with a multi-class Iterative Reweighted Least Squares (IRLS)
algorithm \citep{irls,chen99,krishnapuram,chamroukhiIJCNN2009}.

It can be easily verified that the proposed algorithm is performed
with a time complexity of $O(IJK^3p^2n)$, where $I$ is the number of
iterations of the EM algorithm and $J$ is the average number of
iterations required by its internal IRLS algorithm.

\subsection{Denoising and segmenting a time series}
\label{ssse: estimation of the denoised signal}

In addition to performing time series parametrization, the proposed
approach can be used to denoise and segment time series (or
signals). The denoised time series can be approximated by the
expectation $E(\bx;\hat{\btheta}) =
\big(E(x_1;\hat{\btheta}),\ldots,E(x_n;\hat{\btheta}) \big)$ where
\begin{eqnarray}
E(x_i;\hat{\btheta}) &=& \int_{\IR}x_i p(x_i;\hat{\btheta})dx_i\nonumber\\
            &=&  \sum_{k=1}^K \pi_{ik}(\hat{\bw})\int_{\IR}x_i \mathcal{N}\big(x_{i};\hat{\bbeta}^T_k\bsr_{i},\hat{\sigma}^2_k\big) dx_i\nonumber\\
            &=& \sum_{k=1}^{K} \pi_{ik}(\hat{\bw})\hat{\bbeta}^T_k \bsr_{i} \enspace , \forall i=1,\ldots,n,
\end{eqnarray}
and $\hat{\btheta} = (\hat{\bw},\hat{\bbeta}_1,\ldots,\hat{\bbeta}_K,\hat{\sigma}^2_1,\ldots,\hat{\sigma}^2_K)$ is the parameter vector obtained at convergence of the algorithm. The matrix formulation of the approximated signal $\hat{\bx} = E(\bx;\hat{\btheta})$ is given by:
\begin{equation}
\hat{\bx} = \sum^{K}_{k=1} \hat{\mathbf{\Pi}}_{k} \bT \hat{\bbeta}_{k},
\label{eq. signal expectation}
\end{equation}
where $\hat{\mathbf{\Pi}}_{k}$ is a diagonal matrix whose diagonal elements are the proportions $(\pi_{1k}({\hat{\bw}}),\ldots,\pi_{nk}({\hat{\bw}}))$ associated with the $k^{th}$ regression model. On the other hand, a signal segmentation can also be  obtained by computing the estimated label $\hat{z_i}$ of $x_i$ according to the following rule:
\begin{equation}
\hat{z_i} = \arg \max \limits_{\substack {1\leq k\leq K}} \pi_{ik}(\hat{\bw})\enspace, \quad  \forall i=1,\ldots ,n.
\label{partitionEMLogistique}
\end{equation}
Applying this rule guarantees the time series are segmented into
contiguous segments if the probabilities $\pi_{ik}$ are computed
with a dimension $q=1$ of $\bsw_k$; $k=1,\ldots,K$.
\subsection{Model selection}

In a general application of the proposed model, the optimal values
of $(K,p,q)$ can be computed by using the Bayesian Information
Criterion (BIC) \citep{BIC} which is a penalized
likelihood criterion, defined by
\begin{equation}
\mbox{BIC}(K,p,q) = L(\hat{\btheta};\bx) - \frac{\nu(K,p,q)\log(n)}{2}\enspace,
\end{equation}
where $\nu(K,p,q) = K(p+q+3)-(q+1)$ is the number of parameters of the model and $L(\hat{\btheta};\bx)$ is the log-likelihood obtained at convergence of the EM algorithm. 

\section{Experimental study using simulated signals}
\label{sec: experiments for signals modeling}

This section is devoted to an evaluation of the signal modeling performed by the proposed algorithm using simulated  datasets. For this purpose, the proposed approach was compared with the piecewise regression and the Hidden Markov Regression approaches.

\subsection{Evaluation criteria}
\label{ssec: evaluation criteria}

Two evaluation criteria were used in the simulations. The first criterion is the mean square error between the true simulated curve without noise (which is  the true denoised signal) and the estimated denoised signal given by:
\begin{itemize}
\item $\hat{x}_{i}=\sum_{k=1}^{K} \pi_{ik}(\hat{\bw})\hat{\bbeta}^T_k \bsr_{i}$ for the proposed model;
\item $\hat{x}_{i}=\sum_{k=1}^{K}\hat{z}_{ik}\hat{\bbeta}^T_{k}\bsr_{i}$ for the piecewise polynomial regression model;
\item $\hat{x}_{i}=\sum_{k=1}^{K}\omega_{ik}(\hat{\bPsi})\hat{\bbeta}^T_k \bsr_{i}$ for the HMM regression model.
\end{itemize}
This error criterion is computed by the formula $\frac{1}{n}\sum_{i=1}^{n} [E(x_i;\btheta)-\hat{x}_{i}]^2$. It is used to assess the models with regard to signal denoising and is called  the denoising error.

The second criterion is the misclassification error rate between the simulated and the estimated partitions. It is used to assess the models with regard to signal segmentation. Note that other comparisons between the proposed approach and two versions of the  piecewise  polynomial regression approach including the running time can be found in \citep{chamroukhiIJCNN2009}.

\subsection{Simulation protocol}

The signals were simulated with the proposed regression model with hidden logistic process and all the simulations were performed for a number of segments $K=3$. We chose the value $q=1$ which guarantees a segmentation into contiguous intervals for the proposed model. We considered that all the time series were observed over $5$ seconds with a constant sampling period ($\Delta t=t_i-t_{i-1}$ is constant).

\begin{figure*}[htbp]
\centering
\begin{tabular}{cc}
\includegraphics[width=6cm]{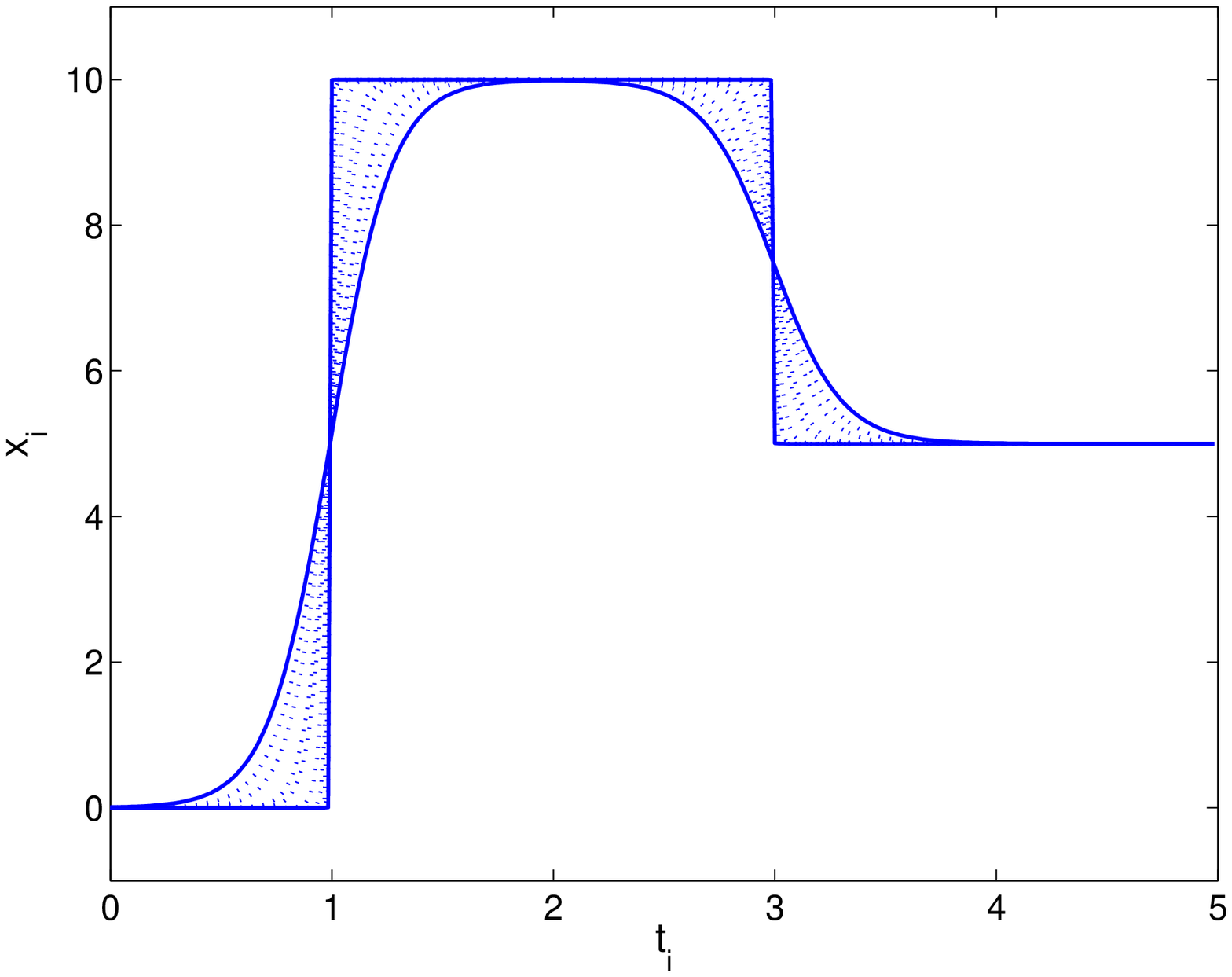}&
\includegraphics[width=6cm]{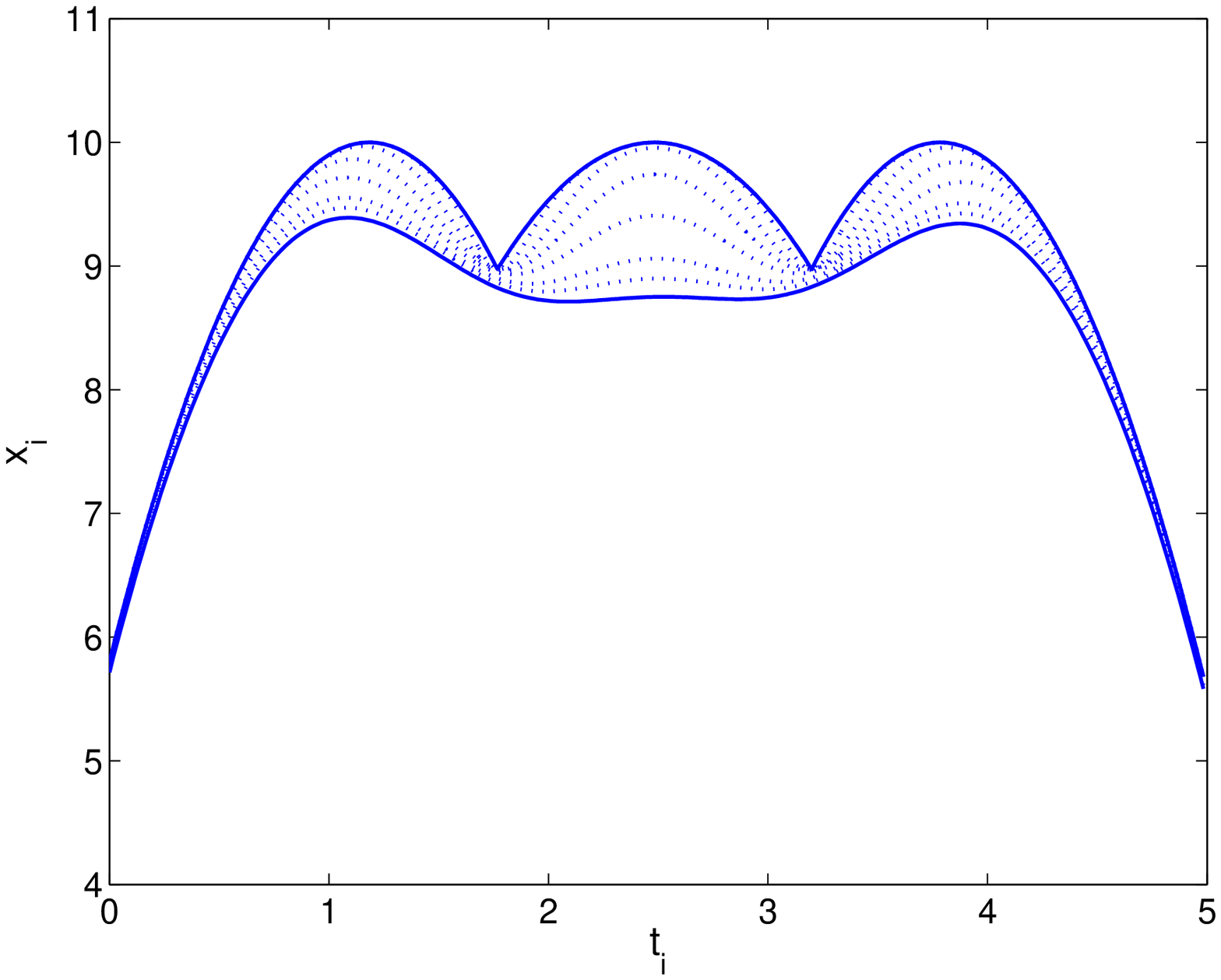}\\
\small{(a)}&\small{(b)}\\
\end{tabular}
\caption{The true denoised signals from abrupt transitions to smooth transitions for situation 1 (a) and situation 2 (b).}
\label{fig.two situations of varynig the smoothess}
\end{figure*}

\begin{figure}[htbp]
\centering
\includegraphics[width=6.1cm]{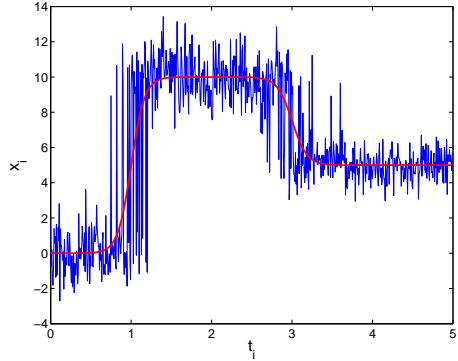}
\caption{Example of simulated signal (with and without noise) for $n=700$ and $\sigma=1$ for situation 1 with a smoothness level of transtion corresponding to the level 8 in Table \ref{table. levels of smoothness}.}
\label{fig. simulated smoothss signal}
\end{figure}

Three experiments were performed: 

\begin{itemize}
\item  the first aims to observe the effect of the smoothness level of transitions on  estimation quality. For this purpose two situations of simulated times series of $n=300$ observations were considered. For the first situation, the time series consisted of three constant polynomial regimes ($K=3, p=0$) with a uniform noise level $\sigma=1$. For the second situation, the time series consisted of three polynomial regimes of order $2$ ($K=3, p=2$) with $n=300$ and $\sigma=0.5$. The set of simulation parameters for the two situations is given in Table \ref{table. parameters of simulation}. The smoothness level of transitions was tuned by means of the term $\lambda_k=\bsw_{k1}; k=1,\ldots,K$, seen in section \ref{ssec: process} and Fig. \ref{logistic_function_k=2_q=1} (a). We used $10$ smoothness levels for each situation. Fig. \ref{fig.two situations of varynig the smoothess} shows the true denoised curves for  situation 1 and situation 2, for the decreasing values of $|\lambda_k|$ shown in Table \ref{table. levels of smoothness}.

\item the second aims to observe the effect of the sample size $n$ on estimation quality. The sample size varied from $100$ to $1000$ is steps of $100$, and the values of the $\sigma_k$ were set to $\sigma_1=1$, $\sigma_2=1.25$, and $\sigma_3=0.75$. Fig. \ref{fig. simulated smoothss signal} shows an example of simulated signal for $n=700$.

\item the third aims to observe the effect of the noise level $\sigma$. The noise level $\sigma$ was assumed to be uniform for all the segments and varied from $0.5$ to $5$ is steps of $0.5$, and the sample size was set to $n=500$.
\end{itemize}

 For each value of $n$, each value of $\sigma$ and each value of the smoothness level of transitions we generated 20 samples and the values of assessment criteria were averaged over the 20 samples.

\begin{table}[htbp]
\centering
\small
\begin{tabular}{|l|ll|}
\hline
Situation 1&$\bbeta_1=0$ &$\bsw_1=[3341.33,-1706.96]$ \\
&$\bbeta_2=10$ & $\bsw_2=[2436.97,-810.07]$ \\
&$\bbeta_3=5$ & $\bsw_3=[0,0]$ \\
 \hline
Situation 2& $\bbeta_1=[-0.64,14.4,-6$]& $\bsw_1=[3767.58,-1510.19]$\\
&$\bbeta_2=[-21.25,25,-5]$ & $\bsw_2=[2468.99,-742.55]$ \\
&$\bbeta_3=[-78.64,45.6,-6]$ & $\bsw_3=[0,0]$ \\
 \hline
\end{tabular}
\caption{Simulation parameters}
\label{table. parameters of simulation}
\end{table}

\begin{table*}[htbp]
\centering
\small
\begin{tabular}{|lllllllllll|}
\hline
\begin{tabular}{l}
Smoothness\\ level of transitions
\end{tabular}
&1 &2 &3 &4 &5 &6 &7 &8 &9 &10\\
\hline
(a) $|\lambda_k|$ divided by: & 1 &2 &5 &10 &20 &40 &50 &80 &100& 125\\
(b) $|\lambda_k|$ divided by: & 1 &10 &50 &100 &150 &200 &250 &275 &300 &400\\
\hline
\end{tabular}
\caption{The different smoothness levels from abrupt transitions to smooth transitons for the situations shown in Fig. \ref{fig.two situations of varynig the smoothess}.}
\label{table. levels of smoothness}
\end{table*}

\subsection{Initialization strategies and stopping rules}
\label{ssec:initializations}

The proposed algorithm and the Hidden Markov regression algorithm were initialized as follows:
\begin{itemize}
\item In the proposed model $\bw$ was set to the null vector;
\item In the HMRM the initial probabilities were set to \linebreak $\pi=(1,0,\ldots,0)$ and $A_{\ell k}=0.5$ for  $\ell \leq k \leq \ell+1$;
\item to initialize $\bbeta_k$ and $\sigma^2_k$, for $k=1,\ldots,K$, several random segmentations of the signal into $K$ segments were used as well as a uniform segmentation. On each segment $k$ we fitted a polynomial regression model and then deduced the valued $\bbeta_k$ and $\sigma^2_k$. The solution providing the highest likelihood was chosen.
\end{itemize}
The two algorithms were stopped when the relative variation of the log-likelihood function between two iterations $|\frac{L^{(m+1)}-L^{(m)}}{L^{(m)}}|$ was below $10^{-6}$ or after $1500$ iterations.

\subsection{Simulation results}
\label{ssec: simulation results}

\begin{figure*}[htbp]
\centering
\begin{tabular}{cc}
\includegraphics[width=6.5cm]{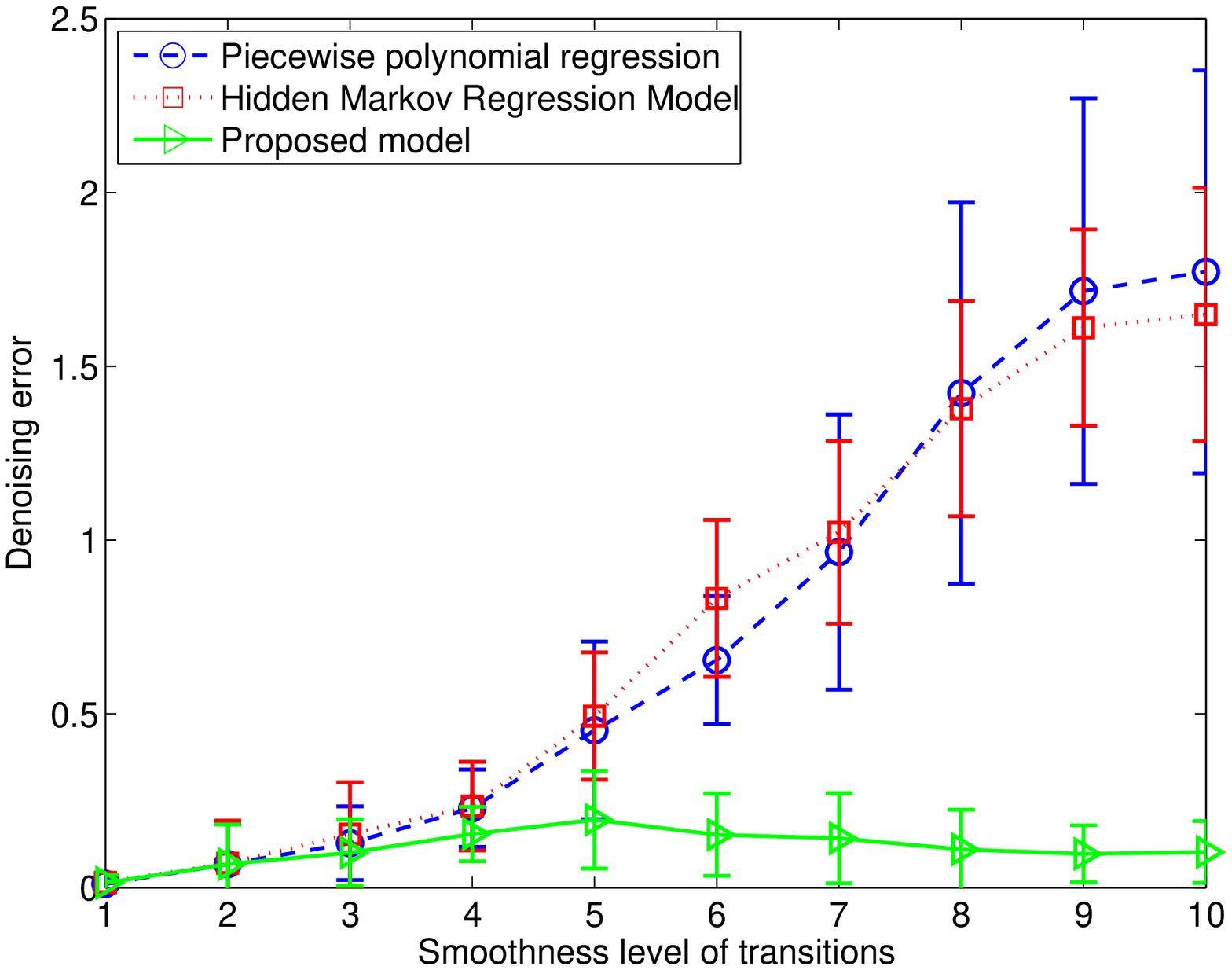}&
\includegraphics[width=6.5cm]{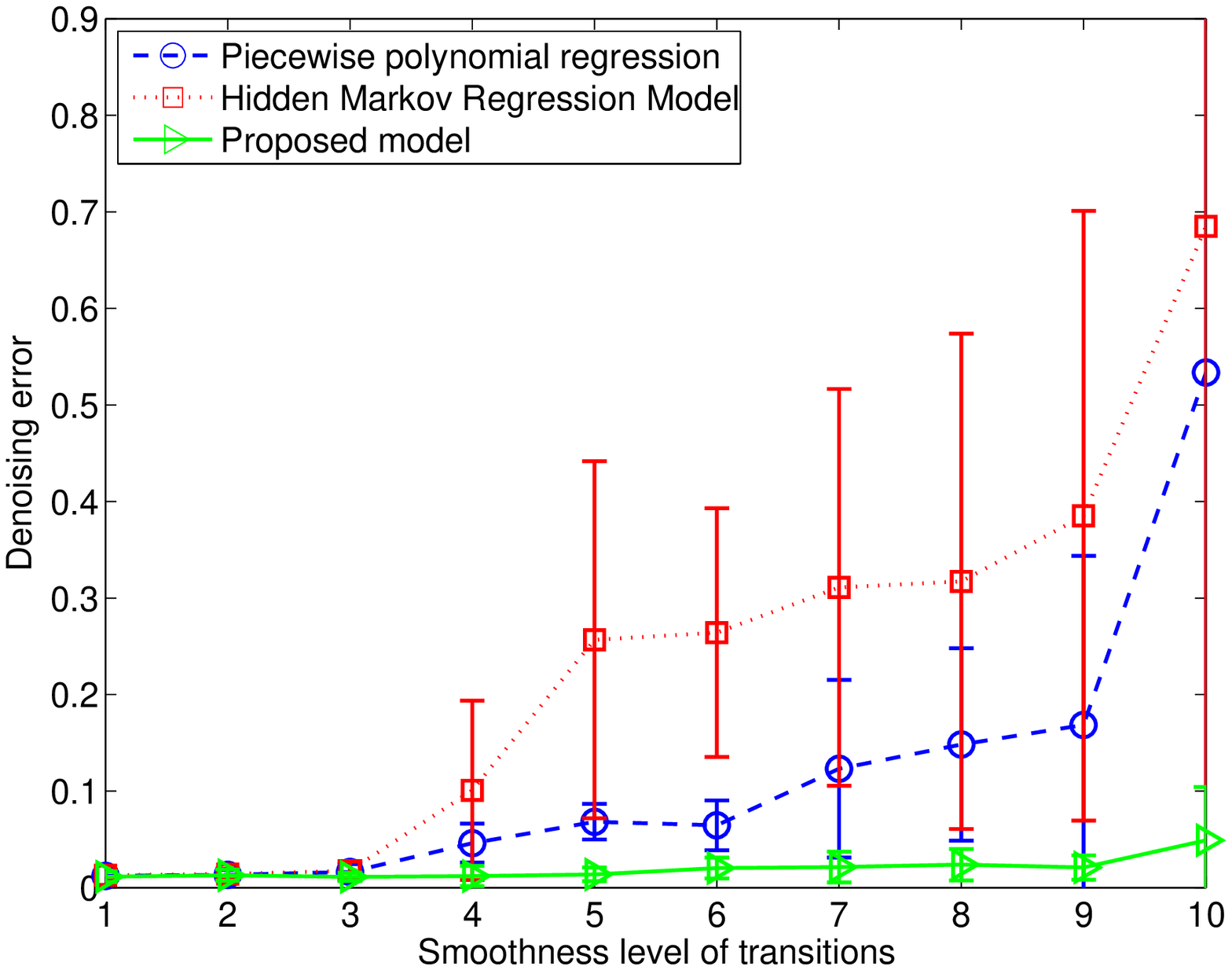}\\
\small{(a)}&\small{(b)}\\
\includegraphics[width=6.5cm]{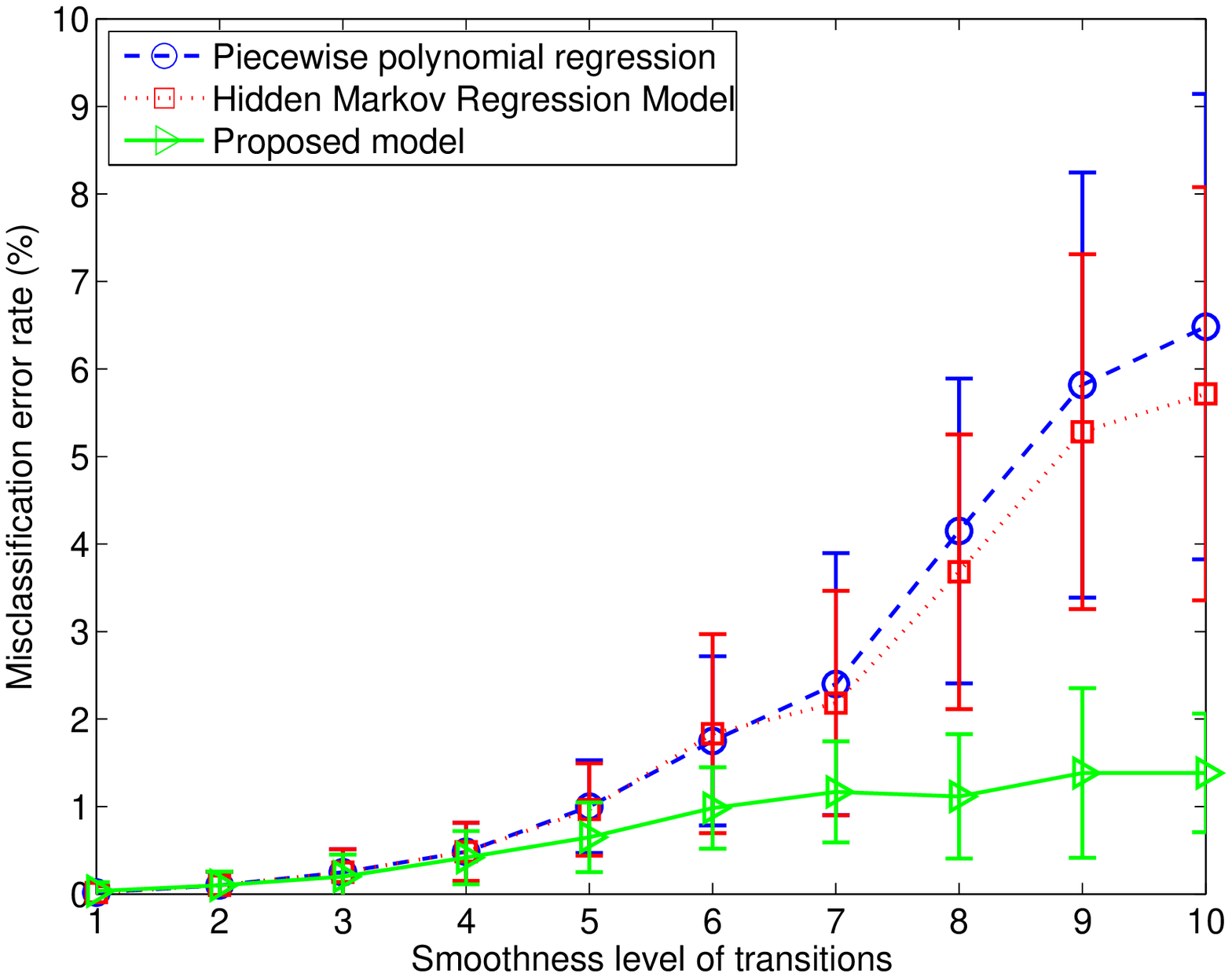}&
\includegraphics[width=6.5cm]{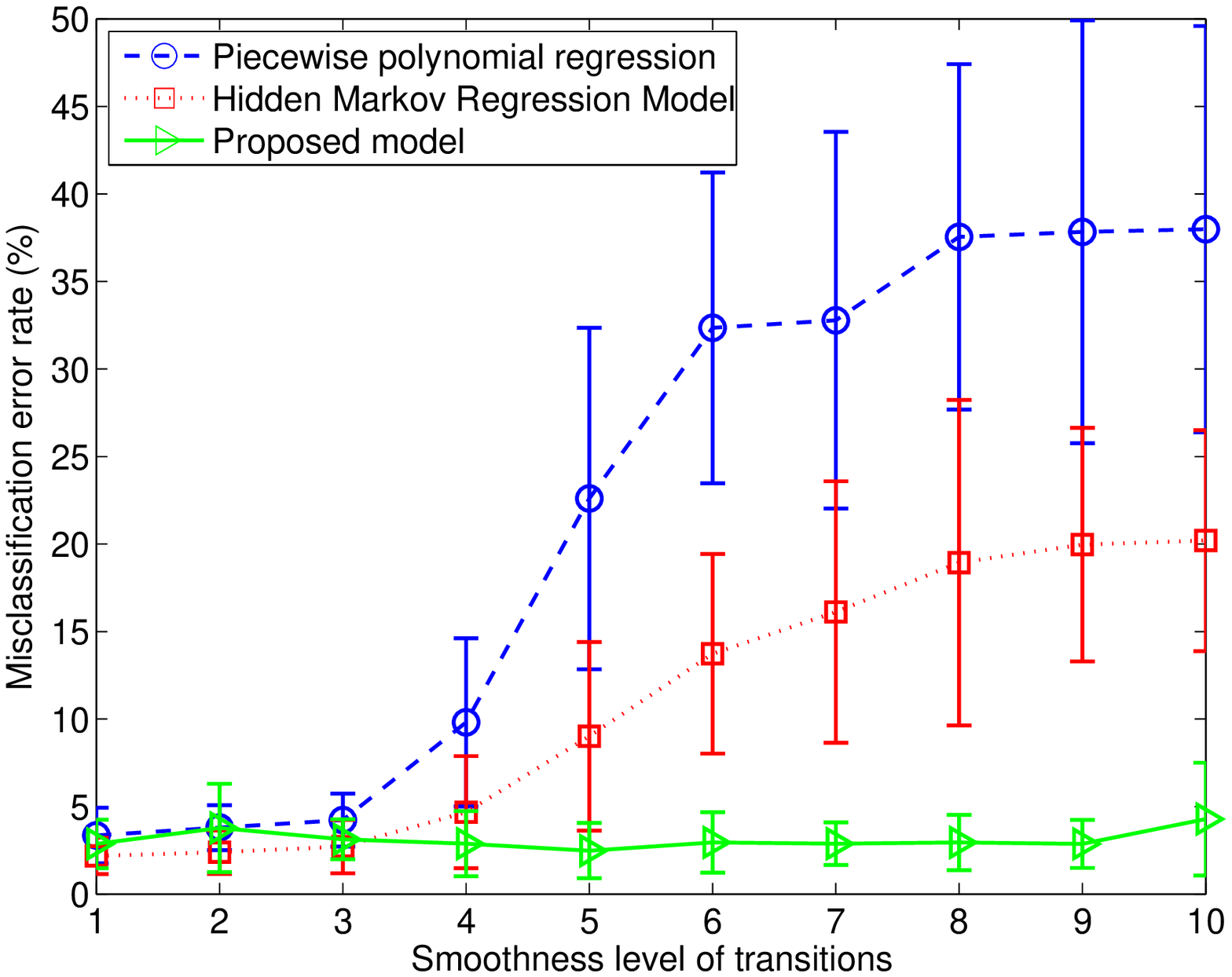}\\
\small{(c)}&\small{(d)}
\end{tabular}
\caption{Denoising error (top) and misclassification error rate (bottom) with the error bars in the range of errors standard deviation, in relation to the smoothness level of transitions,  obtained with the proposed approach (triangle), the piecewise polynomial regression approach (circle) and the HMRM approach (square) for the first situation (left) and for the second situation (right).}
\label{fig. results of varying smoothness}
\end{figure*}

Fig. \ref{fig. results of varying smoothness} shows the denoising error and the misclassification error rate in relation to the smoothness level of transitions  for the first situation (left) and for the second situation (right). It can be seen that the proposed approach performs the signals segmentation and denoising better than the piecewise regression and the HMRM approaches. While the results are closely similar when the transtions are abrupt (until level $3$), the proposed approach provides more accurate results than the two alternatives for smooth transitions for the two situations.

\begin{figure*}[htbp]
\centering
\begin{tabular}{cc}
\includegraphics[width=6.5cm ]{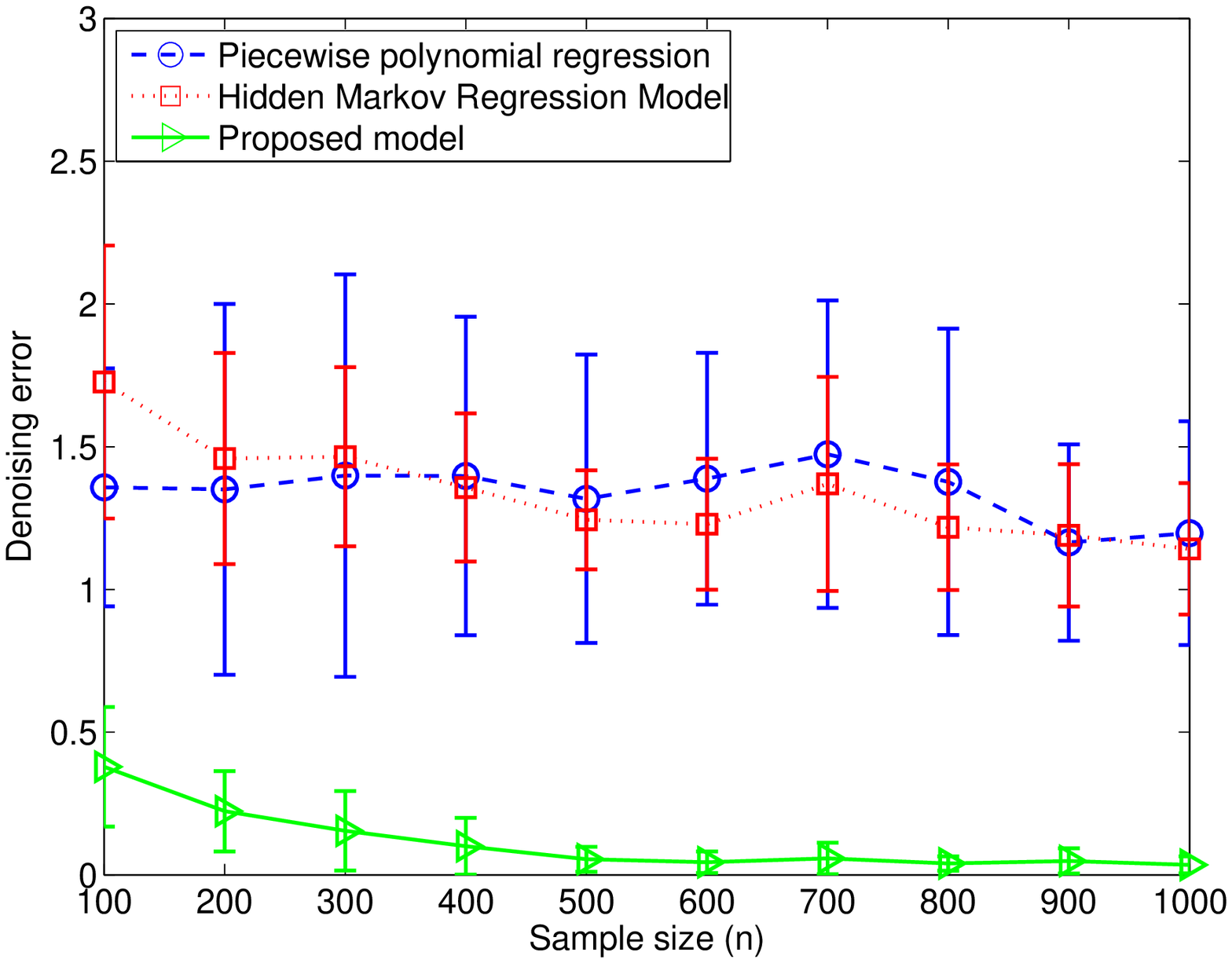}&
\includegraphics[width=6.5cm ]{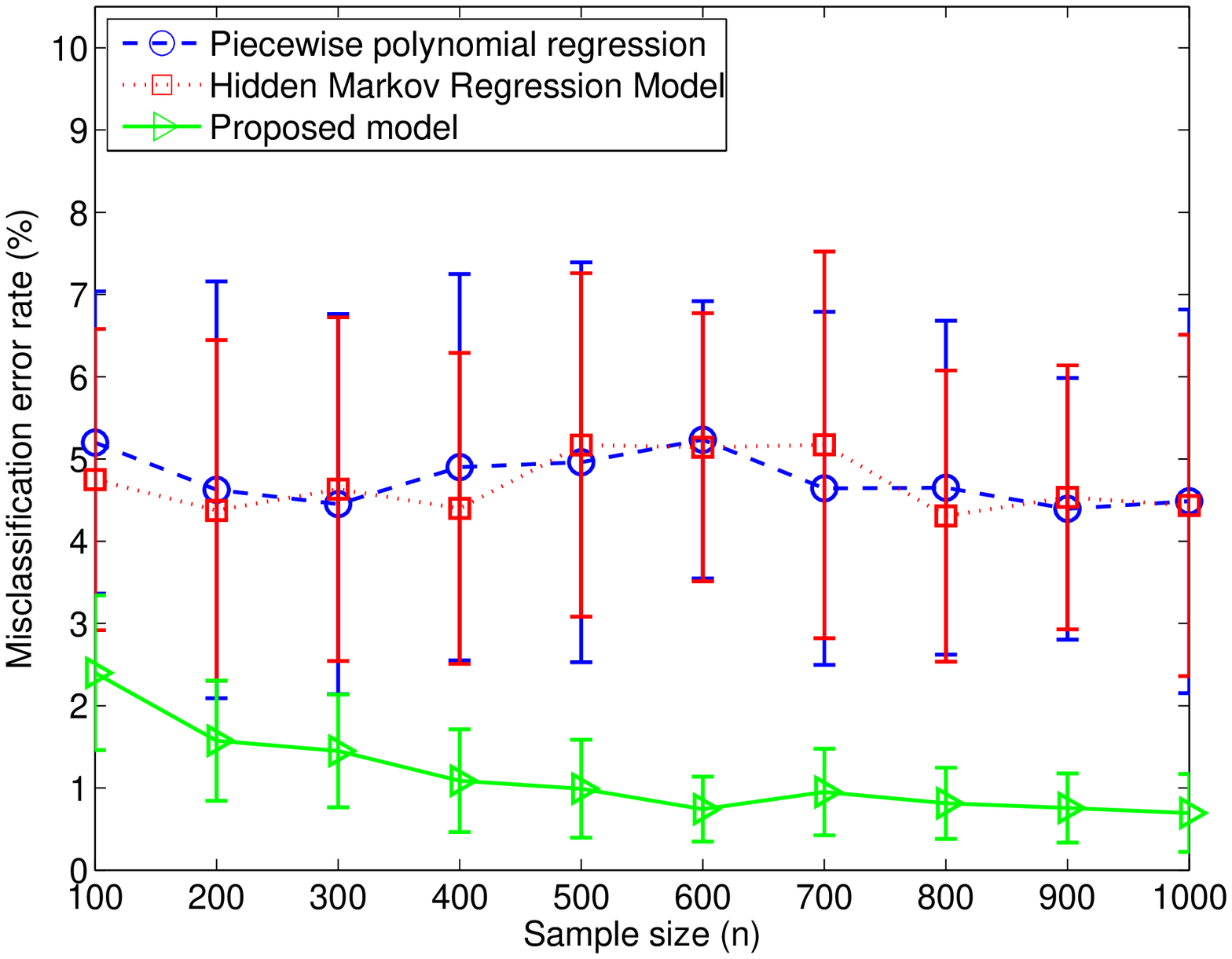}\\
\small{(a)}&\small{(b)}\\
\includegraphics[width=6.5cm ]{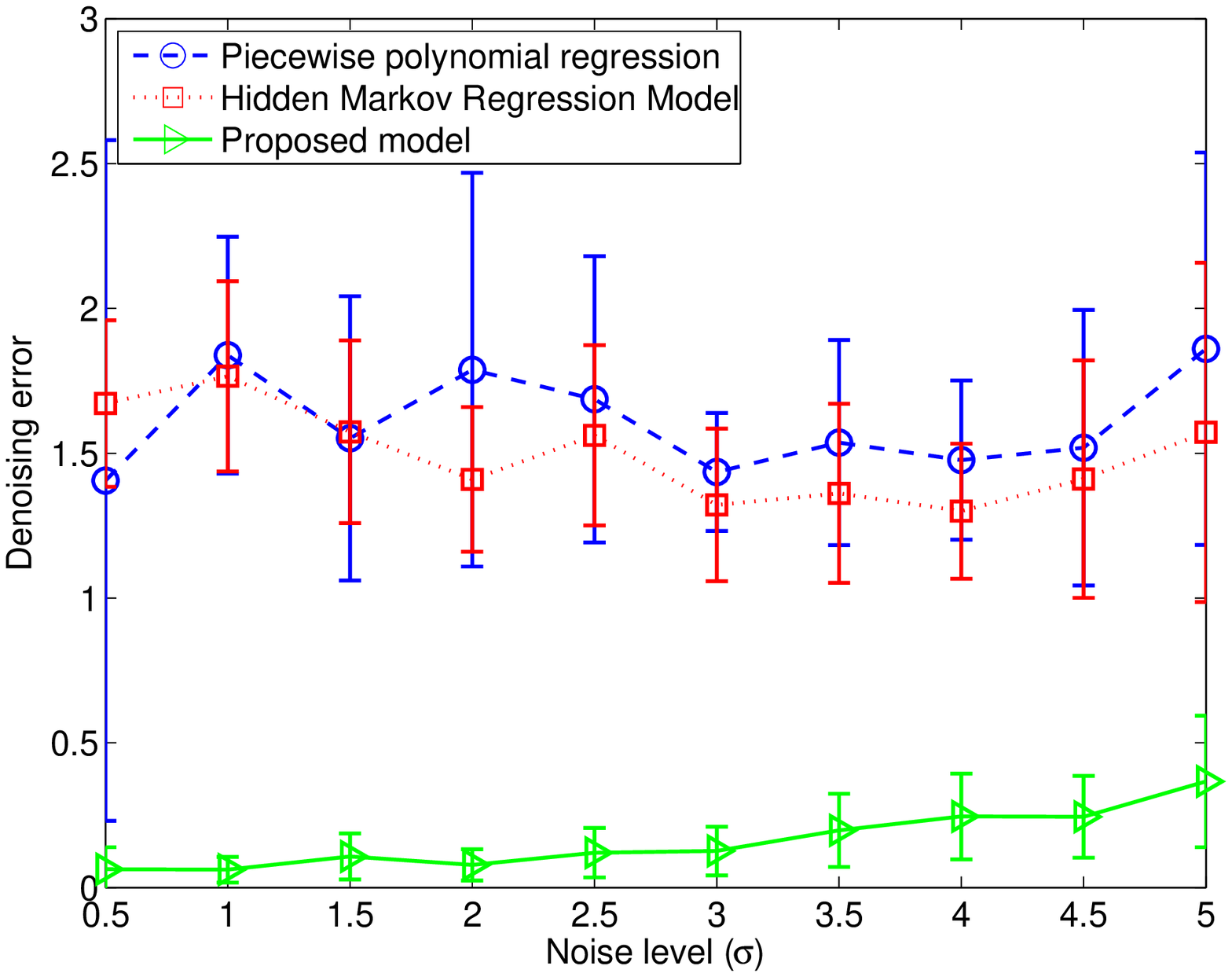}&
\includegraphics[width=6.5cm ]{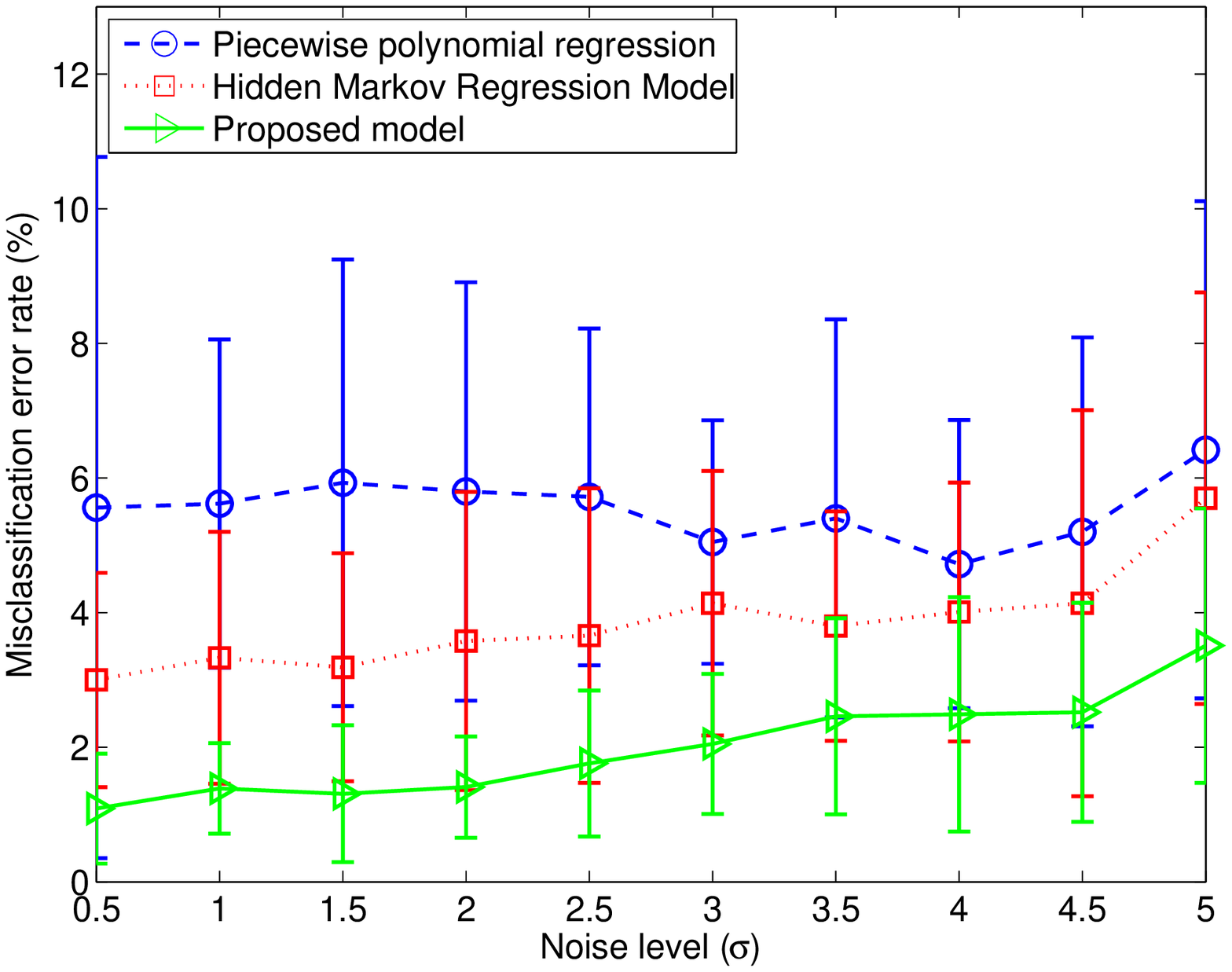}\\
\small{(c)}&\small{(d)}\\
\end{tabular}
\caption{Denoising error (left) and misclassification error rate (right) with the error bars in the range of errors standard deviation, in relation to the sample size $n$ for ($\sigma_1=1, \sigma_2=1.25, \sigma_3=0.75$  (a,b) and the noise level $\sigma$ for $n=500$ (c,d), obtained with the proposed approach (triangle), the piecewise polynomial regression approach (circle) and the HMRM approach (square).}
\label{fig. results_n_sigma_vary}
\end{figure*}

Fig. \ref{fig. results_n_sigma_vary} shows the denoising error and  the misclassification error rate in relation to the sample size $n$ and the noise level $\sigma$.  It can be seen in Fig. \ref{fig. results_n_sigma_vary} (a) and Fig. \ref{fig. results_n_sigma_vary} (b)  that the segmentation error decreases when the sample size $n$ increases for the proposed model which provides more accurate results than the piecewise and the HMRM approaches.   Fig. \ref{fig. results_n_sigma_vary} (c) and Fig. \ref{fig. results_n_sigma_vary} (d) show that when the noise level increases the proposed approach provides more stable  results than to the two other alternative approaches.

\section{Application to real signals}
\label{sec: exp on real data}

This section presents the results obtained by the proposed approach for the switch operation signals modeling and classification. Several types of signals were considered (with and without defects). The number of regression components was chosen in accordance with the number of electromechanical phases of a switch operation ($K = 5$). The value of $q$ was set to $1$, which guarantees segmentation into contiguous intervals for the proposed approach, and the degree of the polynomial regression $p$ was set to $3$   which is appropriate for the different regimes in the signals.

\subsection{Real signal modeling}
\label{ssec: signal modeling}

The proposed regression approach were applied to real signals of switch operations.

\begin{figure*}[htbp]
\centering
\begin{tabular}{cc}
\includegraphics[ width=6.5cm]{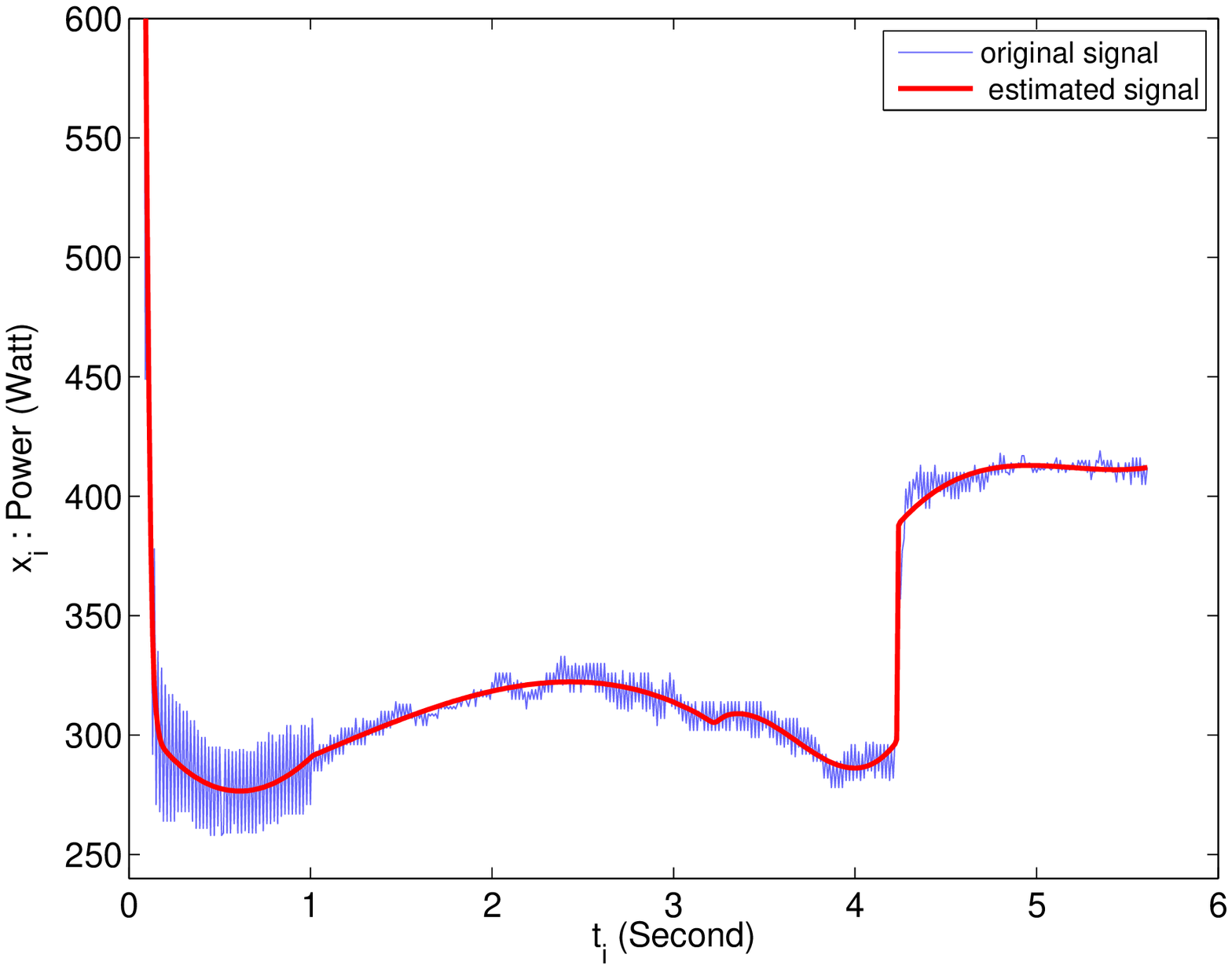}&
\includegraphics[width=6.5cm]{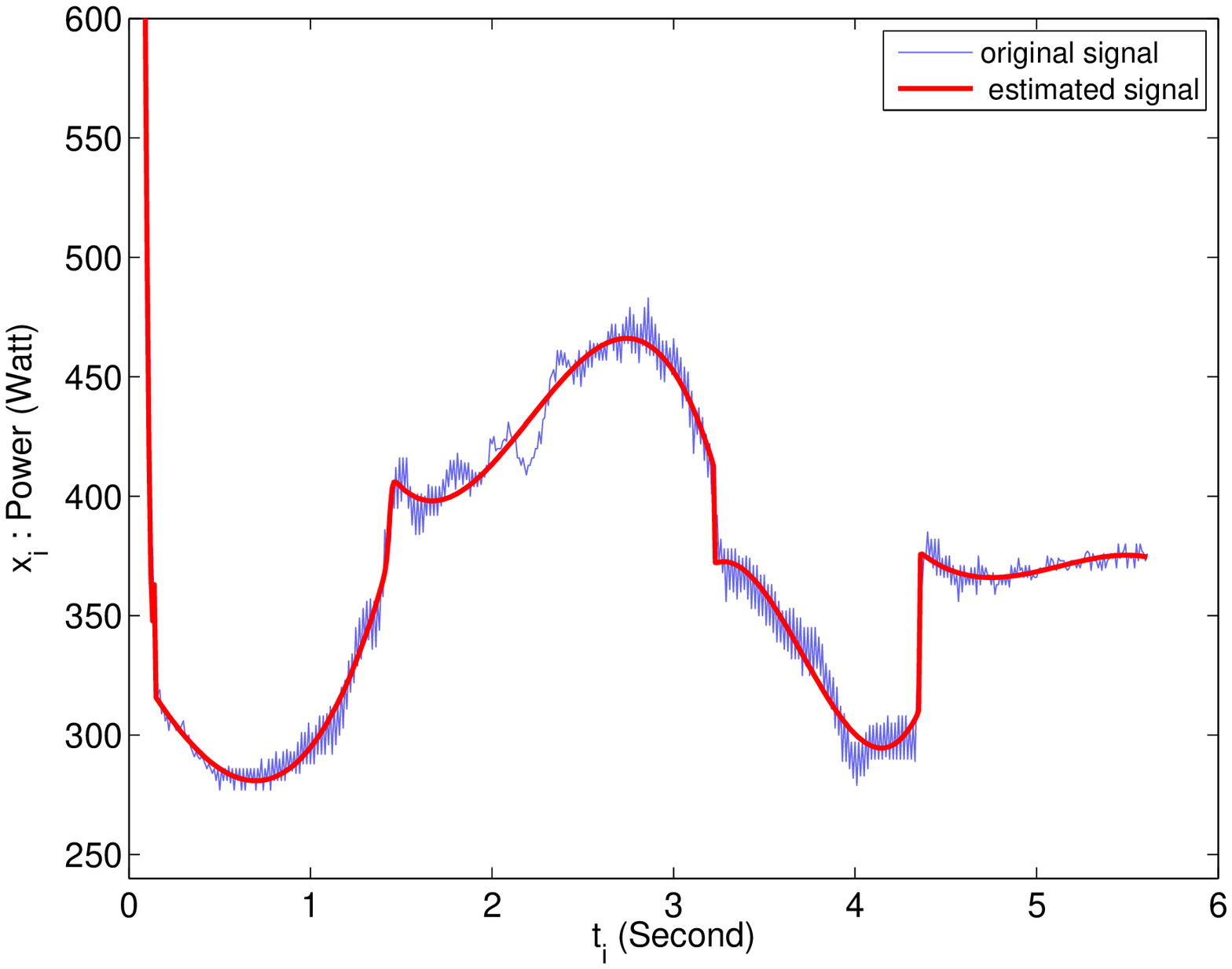} \\
\small{(a)}&\small{(b)}\\
\includegraphics[ width=6.5cm]{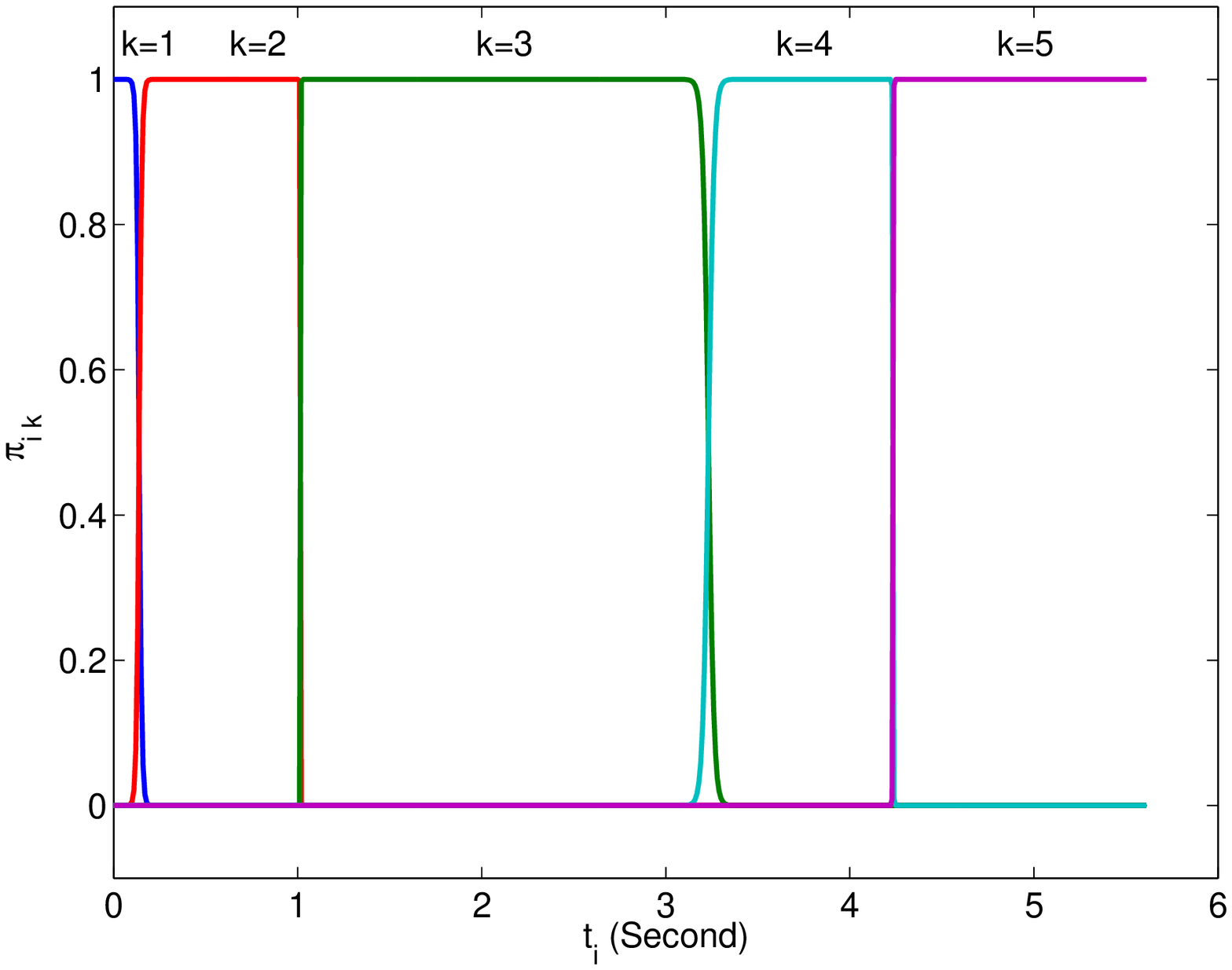} &
\includegraphics[ width=6.5cm]{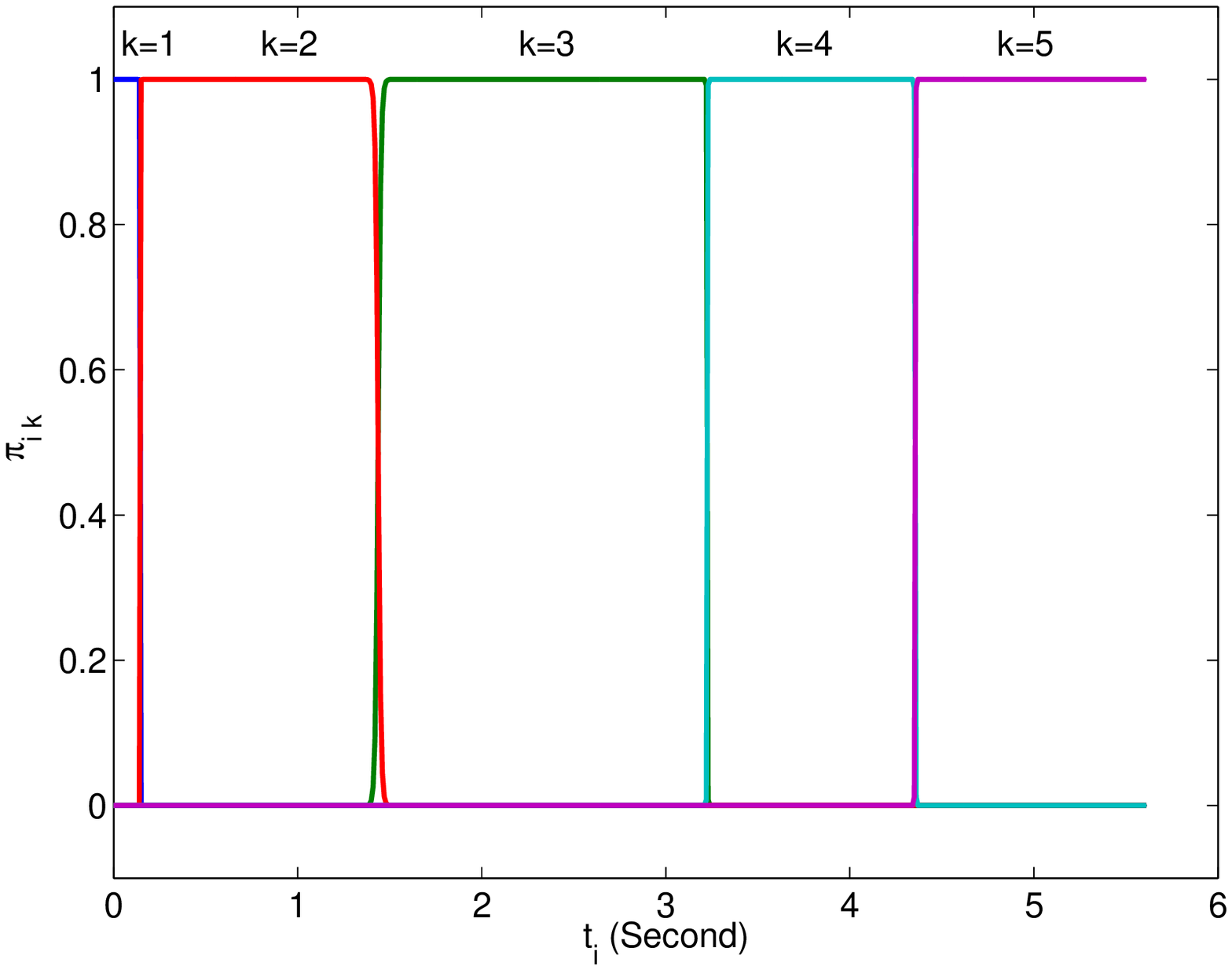}\\
\small{(c)}&\small{(d)}\\
\end{tabular}
\caption{Results obtained with the proposed approach for a signal without defect (a) and a signal with defect (b) with the original signal (in blue) and the estimated signal (in red) and the proportions $\pi_{ik}$, $k=1,\ldots,5$ for the estimated polynomial regression components over time (c) and (d).}
\label{fig. switch_signals_results_proposed_model}
\end{figure*}

Fig. \ref{fig. switch_signals_results_proposed_model} (top) shows the original signals and the denoised signals (the denoised signal provided by the proposed approach is given by equation (\ref{eq. signal expectation})). Fig. \ref{fig. switch_signals_results_proposed_model} (bottom) shows the variation of the probabilities $\pi_{ik}$ over time. It can be seen that these probabilities are very closed to $1$ when the $k^{th}$ regression model seems to be the most faithful to the original signal. 

\subsection{Real signal classification}
\label{ssec: signal classification}

This part is devoted to an evaluation of the classification accuracy of the proposed approach. A database of $N=119$ real signals with known classes was used. This database was divided into two groups: a training base of 84 signals for learning the classes parameters and a test base of 35 signals for evaluating the classifier. The three parametrization methods were applied to applied to all the signals of the database, and the estimated parameters provided by each approach were used as the signal feature vector. After the parametrization step, the MDA was applied to the features extracted from the signals in the training data set. After the learning step, each signal, represented by its feature vector was classified using the Maximum A Posteriori (MAP) rule.

Three different classes of signals indexed by  $g=1,..,3$,  corresponding to the different operating states of the switch mechanism were considered. Thus, the considered classes were
\begin{itemize}
\item $g=1$: no defect class;
\item $g=2$: minor defect class;
\item $g=3$: critical defect class.
\end{itemize}

In what follows we shall use $\bsy_j$ to denote the feature vector $\btheta$ extracted from the signal $\bx_j$, where the index $j =1,\ldots,N$ corresponds to the signal number.

\subsubsection{Modeling the operating classes with mixture models}
\label{ssec: modeling the operating classes with mixture models}

Given a labelled collection of extracted features, the parameters of each class are learned using the Mixture Discriminant Analysis (MDA) \cite{tibshiraniMDA}. In this approach, the density of each class $g=1,..,G$ with $G=3$ is modeled by a Gaussian mixture distribution \cite{tibshiraniMDA,mclachlanFiniteMixtureModels} defined by
\begin{equation}
p(\bsy_j|C_j=g;\bTheta_g) = \sum_{r=1}^{R_g} \alpha_{gr} \mathcal{N} \left(\bsy_j;\bsmu_{gr},\Sigma_{gr}\right),
\end{equation}
where $C_j$ is the discrete variable which takes its value in the set  $\{1,\ldots,3\}$ representing the class of the signal  $\bx_j$, $$\bTheta_g = \left(\alpha_{g1},\ldots,\alpha_{gR_g}, \bsmu_{g1},\ldots,\bsmu_{gR_g},\ldots,\Sigma_{g1},\ldots, \Sigma_{gR_g}\right)$$ is the parameter vector of the mixture density of the class $g$ with $R_g$ is the number of mixture components  and the $\alpha_{gr}$ $(r=1,\ldots,R_g)$ are the mixing proportions satisfying $\sum_{r=1}^{R_g}\alpha_{gr}=1$. The optimal number of Gaussian distributions $R_g$ for each class $g$ is computed by maximizing the BIC criterion \citep{BIC}:
\begin{equation}
\mbox{BIC}(R_g)=L(\hat{\bTheta}_g)-\frac{\nu_{R_g}}{2} \log(n_g),
\end{equation}
where $\hat{\bTheta}_g$  is the maximum likelihood estimate of $\bTheta_g$ provided by the EM algorithm, $\nu_{R_g}$ is the dimension of the parameter vector $\bTheta_g$, and $n_g$ is the cardinal number of  class $g$.

Given the parameter vectors $\hat{\bTheta}_1$, $\hat{\bTheta}_2$, $\hat{\bTheta}_3$ estimated by the EM algorithm for the three classes of signals, each new signal designed by the feature vector $\bsy_j$ is assigned to the class $\hat{g}$ that maximizes the posterior probability that $\bsx_i$ belongs to the class $g$, with respect to $g=1,\ldots,G$:
\begin{equation}
\hat{g}=\arg \max \limits_{\substack{1\leq g\leq G}} p(C_j=g|\bsy_j;\hat{\bTheta}_g),
\label{MAP rule}
\end{equation}
with
\begin{equation}
p(C_j=g|\bsy_j;\hat{\bTheta}_g)=\frac{p(C_j=g)p(\bsy_j|C_j=g;\hat{\bTheta}_g)}{\sum_{g'=1}^{G}p(C_j=g')p(\bsx_i|C_j=g';\hat{\bTheta}_{g'})} ,
\end{equation}
where $p(C_j=g)$ is the prior probability of the class $g$ estimated by the proportion of the signals belonging to class $g$ in the learning phase.

\subsubsection{Classification results}
\label{ssec: classification results}

The results in terms of correct classification rates are given in table (\ref{table. classification results MDA}) and the number of mixture components estimated by the BIC criterion for each class $g$,  for the proposed modeling method,  is given in table (\ref{table. BIC results MDA}).

\begin{table}[htbp]
\centering
\small
\begin{tabular}{|l|c|}
\hline
Modeling approach& \begin{tabular}{c}Correct classification rate (\%)\end{tabular}\\
\hline
Piecewise regression model& \begin{tabular}{c}83\end{tabular}\\
HMRM& \begin{tabular}{c}89 \end{tabular}\\
Proposed regression model& \begin{tabular}{c}91\end{tabular}\\
\hline
\end{tabular}
\caption{Correct classification rates.}
\label{table. classification results MDA}
\end{table}
The correct classification rates  clearly show that using the proposed regression approach for signals modeling outperforms the two alternative approaches.
\begin{table}[htbp]
\centering
\small
\begin{tabular}{|l|ccc|}
\hline
Class $g$ &1 &2 & 3\\
\hline
Number of mixture components $R_g$&1 &1 &3 \\
\hline
\end{tabular}
\caption{Number of mixture components selected with the BIC criterion.}
\label{table. BIC results MDA}
\end{table}
The number of mixture components $R_g=3$ selected with the BIC criterion for the third class (critical defect class) is attributed to the fact that this class contains signals covering a wide range of defects.

\section{Conclusion}
\label{sec: conclusion}
This paper proposes a new approach for time series modeling, in the context of the railway switch mechanism diagnosis. It is based on a regression model incorporating a discrete hidden logistic process. The logistic probability function  used for the hidden variables allows for smooth or abrupt transitions between various polynomial regression components over time. In addition to time series parametrization, the proposed model can provide accurate signal segmentation and denoising. The performance of this approach in terms of signal modeling has been evaluated by comparing it to the piecewise polynomial regression approach and the Hidden Markov  Regression Mode using simulated data and real data. Based on the proposed modeling approach, a mixture discriminant approach has been implemented to classify real signals.

\section*{Acknowlegment} The authors thank the SNCF company and especially M. Antoni from the Infrastructure Department for availability of data.

\bibliographystyle{apalike}
\bibliography{bibliographie_NN_2009}

\end{document}